%% file: PASA.tex
\documentclass[
  journal=pasa,
  manuscript=research-paper,
  year=2023,
  volume=??,
]{cup-journal}

\usepackage{multirow,overpic}
\usepackage{amsmath,amssymb}
\usepackage{upgreek}
\usepackage[nopatch]{microtype}
\usepackage{booktabs}
\usepackage[utf8]{inputenc}

\title{Modelling repetition in {\sc zDM}: a single population of repeating fast radio bursts can explain CHIME data}

\author{C.W.~James}
\affiliation{International Centre for Radio Astronomy Research, Curtin University,Bentley, WA 6102, Australia}
\email[C.W.~James]{clancy.james@curtin.edu.au}


\input{commands.tex}

\begin{document}

\begin{abstract}

Regardless of whether or not all fast radio bursts (FRBs) repeat, those that do form a population with a distribution of rates. This work considers a power-law model of this population, with rate distribution $\Phi_r \sim R^\rr$ between \rmin\ and \rmax. The \zdm\ code is used to model the probability of detecting this population as either apparently once-off or repeat events as a function of redshift, $z$, and dispersion measure, DM. I demonstrate that in the nearby Universe, repeating sources can contribute significantly to the total burst rate. This causes an apparent deficit in the total number of observed sources (once-off and repeaters) relative to the distant Universe that will cause a bias in FRB population models.
Thus instruments with long exposure times should explicitly take repetition into account when fitting the FRB population.

I then fit data from The Canadian Hydrogen Intensity Mapping Experiment (CHIME). The relative number of repeat and apparently once-off FRBs, and their DM, declination, and burst rate distributions, can be well-explained by 50--100\% of CHIME single FRBs being due to repeaters, with $\rmax > 0.75$\,day$^{-1}$ above $10^{39}$\,erg, and $\rr = -2.2_{-0.8}^{+0.6}$. This result is surprisingly consistent with follow-up studies of FRBs detected by the Australian Square Kilometre Array Pathfinder (ASKAP).
Thus the evidence suggests that CHIME and ASKAP view the same repeating FRB population, which is responsible not just for repeating FRBs, but the majority of apparently once-off bursts.

For greater quantitative accuracy, non-Poissonian arrival times, second-order effects in the CHIME response, and a simultaneous fit to the total FRB population parameters, should be treated in more detail in future studies.

\end{abstract}

\input{01_intro.tex}

\input{02_repeater_models.tex}

\input{03_examples.tex}

\input{04_modelling_chime.tex}

\input{05_preliminary_results.tex}

\input{06_fitting_results.tex}

\input{07_future_prospects.tex}

\input{08_systematics}

\input{09_conclusions}

\begin{acknowledgement}
I thank Evan~Keane and J.~Xavier~Prochaska for helpful comments made on the manuscript, Keith Bannister for the motivation for \secref{sec:tdist}, and Casey Law for the motivation for \secref{sec:numbers}.

This research has made use of the NASA/IPAC Extragalactic Database (NED),
which is operated by the Jet Propulsion Laboratory, California Institute of Technology,
under contract with the National Aeronautics and Space Administration. This research made use of Python libraries \textsc{Matplotlib} \citep{Matplotlib2007}, \textsc{NumPy} \citep{Numpy2011}, and \textsc{SciPy} \citep{SciPy2019}. 

This research was partially supported by the Australian Government through the Australian Research Council's Discovery Projects funding scheme (project DP210102103).

\end{acknowledgement}

\paragraph{Funding Statement}

I acknowledge support by the Australian Government through the Australian Research Council's Discovery Projects funding scheme (project DP210102103).

\paragraph{Competing Interests}

None

\paragraph{Data Availability Statement}

The code used is this work is available at \url{https://github.com/FRBs/zdm}.

\printendnotes

\bibliography{PASA}

\appendix

\input{10_appendix}

\end{document}

%% file: commands.tex
\newcommand{\tabref}[1]{Table~\ref{#1}}
\newcommand{\figref}[1]{Figure~\ref{#1}}
\newcommand{\secref}[1]{\S\ref{#1}}

\newcommand{\cat}{\ensuremath{\rm Cat1}}
\newcommand{\gold}{Gold25}
\newcommand{\shin}{S22} 
\newcommand{\jh}{J22c}

\newcommand{\Hnot}{\ensuremath{{H_{0}}}}
\newcommand{\hubbunit}{\ensuremath{\rm km \, s^{-1} \, Mpc^{-1}}}

\newcommand{\pccc}{\ensuremath{{\rm pc}\,{\rm cm}^{-3}}}

\newcommand{\DM}{{\rm DM}}
\newcommand{\dmunits}{\ensuremath{{\rm pc \, cm^{-3}}}}

\newcommand{\dmhost}{\ensuremath{{\rm DM}_{\rm host}}}
\newcommand{\dmmw}{\ensuremath{{\rm DM}_{\rm MW}}}
\newcommand{\dmhalo}{\ensuremath{{\rm DM}_{\rm halo}}}
\newcommand{\dmism}{\ensuremath{{\rm DM}_{\rm ISM}}}
\newcommand{\dmeg}{\ensuremath{{\rm DM}_{\rm EG}}}

\newcommand{\emax}{\ensuremath{E_{\rm max}}}
\newcommand{\Eth}{\ensuremath{E_{\rm th}}}

\newcommand{\eh}{erg\,Hz$^{-1}$}

\newcommand{\tfield}{\ensuremath{T_{\rm f}}}
\newcommand{\rmax}{\ensuremath{R_{\rm max}}}
\newcommand{\rmin}{\ensuremath{R_{\rm min}}}
\newcommand{\rr}{\ensuremath{{\gamma_r}}} 
\newcommand{\rstar}{\ensuremath{R^*}}
\newcommand{\FC}{\ensuremath{F_{\rm single}}}

\newcommand{\zdm}{{\sc zDM}}

\newcommand{\snr}{\ensuremath{{\rm SNR}}}

\newcommand{\muhost}{\ensuremath{\mu_{\rm host}}}
\newcommand{\sigmahost}{\ensuremath{\sigma_{\rm host}}}

\newcommand{\sfrn}{\ensuremath{n_{\rm sfr}}}



%% file: 01_intro.tex
\section{Introduction} \label{sec:intro}

Of the many mysteries surrounding fast radio bursts \citep[FRBs; millisecond duration radio signals arising from cosmological distances][]{Lorimer2007,Thornton2013} the question of whether or not they all repeat remains one of the greatest \citep[e.g.][]{Caleb2019repeaters}. Since the discovery of the first repeating FRB, FRB\,20121102A \citep{spitler2016}, its localisation to a dwarf galaxy and association with a persistent radio source \citep[PRS;][]{Chatterjee2017}, combined with its complex time--frequency burst structure \citep{Hessels2019}, have distinguished it from populations of apparently once-off FRBs, which arise from a plethora of galaxy types \citep{Bhandari2020a,Heintz2020,Bhandari+22,Gordon+2023}, and are more likely to exhibit broadband, single-component morphologies \citep{CHIME_morphology_2021}. Further discoveries of repeating FRBs have somewhat muddled this simple dichotomy however --- while FRB~20190520B is similar in both its bursts and its host galaxy to FRB~20121102A \citep{Niu2022}, FRB~20180916B is located in a large spiral galaxy, close to --- but offset from --- a star-forming region \citep{MarcoteRepeaterLocalisation2020}, while FRB~20200120E arises from a globular cluster \citep{CHIME_M81_2021,Kirsten2022_GC}, which as is typical for such clusters, has no apparent star-forming activity.

Models of the FRB population paint a similarly ambiguous picture. Very early FRB data was shown to be consistent with all FRBs being similar to FRB~20121102A \citep{Lu2016_EDF}. As more data became available, \citet{Caleb2019repeaters} was able to rule out that all FRBs could repeat as rapidly as FRB~20121102A, while \citet{James2019b_pop_limits} showed that the number of strong repeaters similar to FRB\,20121102A must be at most 27\,Gpc$^{-3}$ with 90\% confidence. Strong limits on the repetition rate of individual FRBs now demonstrate that at least some repeat at most very rarely \citep{James2020a_followup,Lin2023_CHIME_far_sidelobe}.
More recently, \citet{Gardinier2021_frbpoppy_repeaters} used {\sc FRBPOPPY} \citep{Gardenier2019FRBPOPPY} to match a population of repeating FRBs to the dispersion measure (DM) distribution from the Canadian Hydrogen
Intensity Mapping Experiment \citep[CHIME;][]{CHIME2018_system}. They confirm that for a given population of repeating FRBs, those in the nearby Universe will preferentially be detected as repeaters, while those in the distant Universe will more likely be viewed as once-off bursts. This effect is qualitatively present in the CHIME data: repeaters at low DM have a higher rate \citep[as noted by D.\ Good at the FRB~2021 online conference, using data from ][]{CHIME_catalog1_2021}, and the mean DM of repeaters is lower than that of once-off bursts \citep{CHIME_2023_25reps}. However, no quantitative analysis has attempted to fit this effect, which must be present in the data. Thus the degree to which this is evidence for repeating FRBs and intrinsically once-off bursts being distinct populations remains unknown.

Another result from FRB population analysis is the all-burst luminosity function, which is usually modelled as either a power-law or a Schechter function \citep[e.g.][]{Luo2020,James2022Meth,Shin2022}. This is generally fitted to have a comparable slope to original measurements of the FRB~20121102A burst energy distribution \citep{Law2017}, indicating that the all-burst spectrum could be built up as the sum of repeating FRBs. More-detailed measurements of burst energy spectra of repeaters has shown behaviour much more complex than a single power-law however \citep{Li2021_FAST_121102,Hewitt2021,Zhang2023FAST20220912A,FAST2022_active_episode_2}, making comparisons difficult, though there are indications that the very high-energy tail might remain consistent with results from population models \citep{Kirsten2023_highenergy}. Again, the evidence for or against an FRB population dominated by intrinsically repeating FRBs or once-off events is ambiguous.

Results from both host galaxy and population modelling therefore indicate that FRB progenitors either come from multiple populations, or if they do all repeat, must come from a broad distribution of repetition rates. The different morphologies observed for repeating and once-off bursts is consistent with both pictures \citep{CHIME_morphology_2021}; as is the observation of PRSs associated with some of the brightest and most rapid repeating objects \citep{Chatterjee2017,Niu2022}, since it is not implausible PRSs dissipate and pulse morphologies change as the progenitor ages. Even if some FRBs arise from intrinsically cataclysmic events, such as black hole formation following a binary neutron star merger \citep{Zhang2014,Falcke,Moroianu2023}, it should be expected that the portion of the population that does intrinsically repeat has a broad distribution of properties. And from an observational or modelling perspective, there is no practical difference between FRBs which repeat very rarely, and those that are intrinsically once-off. This motivates studies which attempt to fit the properties of an intrinsically repeating FRB population to observational data.

FRB observations however give a biased picture of the true underlying FRB population. This is particularly the case with repeaters, which make better targets for host galaxy identification --- of the 492 distinct FRBs published by \citet{CHIME_catalog1_2021}, four repeaters have been localised to their host galaxies \citep{MarcoteRepeaterLocalisation2020,Bhardwaj_NGC3252,CHIME2022_rep_loc}, with a further single burst in that catalog since identified as a repeater and localised to its host \citep{Ibik2023CHIMEhosts}. However, only a single once-off CHIME FRB has a tentative host galaxy identification \citep{Panther2022_GWFRBhost}. Furthermore, once trends in FRB behaviour become identified, these result in preferential targeting with follow-up observations, e.g.\ the identification of FRB~20180301A as a repeater from its time--frequency structure \citep{Price2019_breakthrough,Luo_diverse_polarisation}, or the targeting of FRB~20121102A during its identified activity phase \citep{2020_121102_periodicity1,Li2021_FAST_121102}. Thus once a trend is identified, follow-up observations will naturally reinforce them. It is important to note that CHIME detects repeat bursts from FRBs in an almost unbiased manner \citep{CHIME2019c_eight,CHIME2020a_nine_reps} --- 
the only other repeating FRB to be identified as such is FRB~20190520B \citep{Niu2022}, while
all other identifications have been through targeted follow-up observations.

Observational biases thus limit the use of repeating FRBs in FRB population analyses, which model their cosmological source evolution, burst energy distribution, spectral properties, and local/cosmological/host DM contributions \citep[e.g.][]{Luo2020,James2022Meth}. When they are included, only the first burst of a repeating FRB tends to be used \citep[e.g.][]{Shin2022}, in order to make the analysis as insensitive to their repeating nature as possible. 

We are only aware of one work which fits the intrinsic properties of the repeating FRB population. \citet{James2020b_popreps} use results from follow-up observations of CRAFT FRBs with the Murriyang (Parkes) and Robert C.\ Byrd Green Bank Telescope (GBT) \citep{James2020a_followup}, and assuming a power-law distribution of FRB repetition rates $R$, $dN(R)/dR \propto R^{\rr}$, the authors placed limits on the power-law index $\rr < -2$, and the maximum repetition rate \rmax. The data fit however incorporated only one observed repeater, FRB~20171019A \citep{Shannonetal2018,Kumar2019}, and 19 non-repeaters, in particular FRB~20171020A \citep{Mahony2018}, the proximity of which rules out repetition rates greater than 0.011 day$^{-1}$ above $10^{39}$\,erg \citep{James2020a_followup,Lee-Waddell_2023}. Another work, \citet{Law2022_PRS}, uses CHIME data to model the mean apparent (i.e.\ not intrinsic) repetition rate, finding of 25--440\,yr$^{-1}$. The lower limit was calculated by assuming that only sources observed to repeat were true repeaters, while the upper limit assumed singly observed FRBs were also due to repeaters. Interestingly, the authors note the potential of using the population of PRS to constrain the repeating FRB population, and vice versa.

The aim of this paper is to incorporate a model for repeating FRBs into the framework of the \zdm\ code \citep{zdm,James2022Meth}, as described in \secref{sec:modelling}. Example FRB populations are then used to estimate the biasing effects of FRB repetition on the z--DM distribution of the FRB population in \secref{sec:qualitative}. In \secref{sec:CHIME}, a model of CHIME is described, and in \secref{sec:preliminary}, it's shown how the model compares to CHIME FRBs from Catalogue 1 \citep{CHIME_catalog1_2021}. In \secref{sec:results}, a fit is performed of FRB repetition parameters to CHIME data, and use the results to make predictions for future experiments in \secref{sec:future_prospects}. Potential systematic effects are discussed in \secref{sec:systematics}. The results are compared to those from FRB follow-up observations with the Australian Square Kilometre Array Pathfinder (ASKAP), and estimates of the number density of PRS, in \secref{sec:askap}; findings are summarised in \secref{sec:conclusions}.

Throughout, I use the nomenclature of ``burst'' to refer to a single FRB (which nonetheless may have multiple components, typically on ms or sub-ms scales); while ``FRB'' or ``progenitor'' refers to the progenitor objects, such that a single repeating FRB emits multiple bursts.

A standard Planck cosmology \citep{PlanckCosmology2018} with $\Hnot=67.4$\,\hubbunit\ is used throughout. The Kolmogorov-Smirnov \citep[`KS';][]{kolmogorov,smirnov} test is used to assess consistency between observed and expected distributions. This is a somewhat arbitrary choice, with no strong preference compared to e.g.\ the Anderson-Darling test \citep['AD';][]{AndersonDarling}, though the extra sensitivity of the AD-test to the tails of the distributions analysed herein may not be desirable.

%% file: 02_repeater_models.tex
\section{Modelling repeating FRBs}
\label{sec:modelling}

I begin by characterising repeating FRBs solely by their time-averaged burst rate $R$, which is defined as their intrinsic rest-frame rate of producing bursts above $10^{39}$\,erg.
All FRBs are treated as repeating according to a Poissonian distribution (see \secref{sec:nonpoissonian} for a discussion of this assumption), such that the probability of them producing $N$ observed bursts given an expectation $\lambda = R_{\rm obs} T$ (for some time interval $T$) is
\begin{eqnarray}
P(N) & = & \frac{\lambda^N \exp(-\lambda)}{N!}. \label{eq:poisson}
\end{eqnarray}
It is assumed that each repeating FRB has an identical energy distribution, with cumulative probability described by an upper incomplete Gamma (Schechter) function. Thus the observable rate $R_{\rm obs}$ above some energy $\Eth$ (itself a function of $z$, DM, and position in the telescope beam) scales as
\begin{eqnarray}
R_{\rm obs}(\Eth) = \frac{R}{1+z} \frac{\int_{\Eth}^{\infty} (E/\emax)^\gamma e^{-E/\emax} dE}{\int_{10^{39}}^{\infty} (E/\emax)^\gamma e^{-E/\emax} dE}, \label{eq:robs}
\end{eqnarray}
where $\gamma$ is the cumulative power-law index, and \emax\ some cut-off energy.

The population density of repeating FRBs, $\Phi_r$ (progenitors Mpc$^{-3}$), is modelled with repetition rates above some rate $R$ via a power-law distribution of their intrinsic repetition rates $R$,
\begin{eqnarray}
  & C_r & (R < \rmin) \label{eq:integral_rate_distribution}\\
\Phi_r(R)  = & 0 & (R > \rmax) \nonumber \\
&  C_r \frac{ \left( \frac{R}{\rmin}\right)^{\rr+1} - \left(\frac{\rmax}{\rmin}\right)^{\rr+1}  }{1-\left(\frac{\rmax}{\rmin}\right)^{\rr+1} } &{\rm otherwise}, \nonumber
\end{eqnarray}
between minimum and maximum rates $R_{\rm min}$ and $R_{\rm max}$ respectively, with differential power-law index \rr, and constant population density $C_r$. The differential rate is more commonly used, which is defined as
\begin{eqnarray}
\frac{d\Phi_r(R)}{dR} & = & C^{\prime}_r R^{\rr} \label{eq:rdiff} \\
C^{\prime}_r & = & \frac{(\rr+1) C_r}{\rmin^{\rr+1}-\rmax^{\rr+1}}. \nonumber
\end{eqnarray}
In this model, the total burst density above $10^{39}$\,erg, $C$ (bursts Mpc$^{-3}$ yr$^{-1}$), is related to the population of repeaters via
\begin{eqnarray}
C & = &  \int_{\rmin}^{\rmax} R C^{\prime}_r R^{\rr} dR \nonumber \\
& = & \frac{C^{\prime}_r}{\rr+2} \left[\rmax^{\rr+2} -\rmin^{\rr+2} \right].
\end{eqnarray}
As per previous work, both $C$ and $C_r$ are defined as the number of bursts and FRB progenitors respectively observed at 1.3\,GHz at $z=0$. Both are treated as evolving with the star-formation rate, such that $C_r$ evolves as
\begin{eqnarray}
C_r(z) & = & C_r \left( \frac{{\rm SFR}(z)}{{\rm SFR}(z=0)} \right)^{\sfrn}. \label{eq:sfr}
\end{eqnarray}
\sfrn\ is used to scale smoothly between a non-evolving population ($\sfrn=0$), one evolving with the star-formation rate ($\sfrn=1$), and a population with a stronger peak at high redshift, such as AGN activity ($\sfrn > 1$). In the case of $C$, an additional factor of $(1+z)^{-1}$ is included on the right-hand-side of \eqref{eq:sfr}, due to the time-dilation effect. However, the number of FRB progenitors is unaffected by time-dilation; rather, the time dilation effect is instead included when scaling between the rate seen by an observer and the intrinsic rate.

\subsection{Implementation in the \zdm\ code}

The \zdm\ code was originally developed for the analysis of \citet{James2022Meth}, and has since been extended as per \citet{James2022_H0} and 
\citet{Baptista2023}. It calculates the expected number of FRBs from an FRB survey as a function of their redshift $z$, dispersion measure DM, and relative fluence $F$ compared to threshold fluence $F_{\rm th}$.
The key parameters of the code, and their current best-fit values, are given in \tabref{tab:extreme_params}.

For this analysis, only the measured values of $z$ and DM for an FRB are considered --- unlike previous works, where signal-to-noise ratio (\snr) is also used. I add an additional observable: whether or not an FRB is observed as a repeater. This is treated as a boolean, i.e.\ the number of observed repeats is not used. Furthermore, only repetition information obtained through initial blind surveys is considered, i.e.\ this analysis is not suited to modelling FRBs determined to repeat or not through targeted follow-up observations.
While CHIME remains the only FRB instrument to observe repeat bursts by this definition, the non-observation of repeaters in other blind surveys can still be included.

The total number of repeaters in any given z--DM bin is given by:
\begin{eqnarray}
N(z,{\rm DM}) & = & C_r(z) \frac{dV}{dz d\Omega} \Delta z f_z({\rm DM}) \Delta {\rm DM} \Delta \Omega,
\end{eqnarray}
where $\Delta \Omega$ is the solid angle observed by the beam, $\Delta z$ and $\Delta$DM are the bin sizes of the z--DM grid, $dV (dz d\Omega)^{-1}$ is the size of the cosmological volume element, and $f_z({\rm DM})$ is the distribution of dispersion measures of FRBs at that redshift \citep[itself a function of cosmological and host galaxy contributions, as given in ][]{James2022Meth}.

The distribution of intrinsic FRB rates within that volume element is given by \eqref{eq:rdiff}, which for any given threshold energy \Eth, produces a distribution in $R_{\rm obs}$ as per \eqref{eq:robs}. Thus the expected number of once-off FRBs from that volume becomes
\begin{eqnarray}
&\left<N_1(z,{\rm DM})\right> =  
\frac{dV}{dz d\Omega} \Delta z f_z({\rm DM}) \Delta {\rm DM} \Delta \Omega \label{eq:p1} \\
 &\cdot \int_{\rmin}^{\rmax} R_{\rm obs} T_{\rm obs} e^{-R_{\rm obs} T_{\rm obs}} 
 \frac{d \Phi_r(R)}{d R}
 \frac{dR}{dR_{\rm obs}} dR_{\rm obs}. \nonumber
\end{eqnarray}
Here, I have left off the dependence of $R$ on $R_{\rm obs}$, expressed through \eqref{eq:robs}. Similarly, 
\begin{eqnarray}
&\left<N_0(z,{\rm DM})\right>  =  
\frac{dV}{dz d\Omega}\Delta z f_z({\rm DM}) \Delta {\rm DM} \Delta \Omega \label{eq:p0}\\
&\cdot \int_{\rmin}^{\rmax} e^{-R_{\rm obs} T_{\rm obs}} 
 \frac{d \Phi_r(R)}{d R} 
\frac{dR}{dR_{\rm obs}} dR_{\rm obs} \nonumber
\end{eqnarray}
calculates the number of progenitors for which no bursts are detected. Thus, the expected number of repeating FRBs can be deduced as
\begin{eqnarray}
\left<N_{\rm reps}\right> & = & N - \left<N_1\right> - \left<N_0\right>.
\end{eqnarray}
This therefore allows a likelihood to be assigned to both once-off and repeating FRBs as a function of redshift $z$ and dispersion measure DM.

%% file: 03_examples.tex
\section{The effects of repetition on FRB redshift distributions}
\label{sec:qualitative}

To qualitatively illustrate how the presence of FRB repetition affects the measured FRB redshift distribution, I perform example calculations for the CRAFT incoherent sum \citep[ICS;][]{Bannister2019} survey parameters as described in \citet{James2022Meth}. Since this observing mode is mostly commensal, total time per field \tfield\ to the end of 2022 has been at most $\sim$37 days (Shannon et al., in prep.). For illustrative purposes however, \tfield\ is varied between 10, 100, and 1000 days, and a central frequency of 1.3\,GHz is used.

For repeating FRB parameters, I consider two scenarios. The first is strong repeaters only, setting $\rmax = \rmin = 4$\,day$^{-1}$ (thus making \rr\ irrelevant), which is approximately the time-averaged rate observed above $10^{39}$\,erg for FRB~20121102A by \citet{Li2021_FAST_121102}, divided by four to account for the off part of the activity cycle \citep{2020_121102_periodicity1}; this is also consistent with the time-averaged rate observed by \citet{Law2017}. The second is a more realistic scenario, with a distribution of repeaters with $\rmax=10$\,day$^{-1}$ (i.e.\ slightly above FRB~20121102A), $\rmin=10^{-3}$\,day (slightly below FRB~20171020A), and an index of $\rr=-2.2$, which is consistent with the results of \citet{James2020b_popreps}.

For the remaining properties of the FRB population, note that most previous models  have analysed only once-off FRBs, or alternatively, included only the first burst of repeating FRBs. Several authors have derived values constraining \eqref{eq:robs}, typically obtaining a cumulative fluence index $\gamma \sim -1$, $E_{\rm max} \sim 10^{41-42}$\,erg, with $\mathcal{O} \sim 10^5$ FRBs Mpc$^{-3}$ yr$^{-1}$ in the local Universe, and finding no evidence for a minimum energy \citep{Luo2020,James2022Lett,James2022_H0,Shin2022}. For now, the best-fit values from \citet{James2022_H0} --- given in \tabref{tab:extreme_params} --- are used, and other possibilities are considered in \secref{sec:preliminary}.

\begin{figure}
    \centering
    \includegraphics[width=\columnwidth]{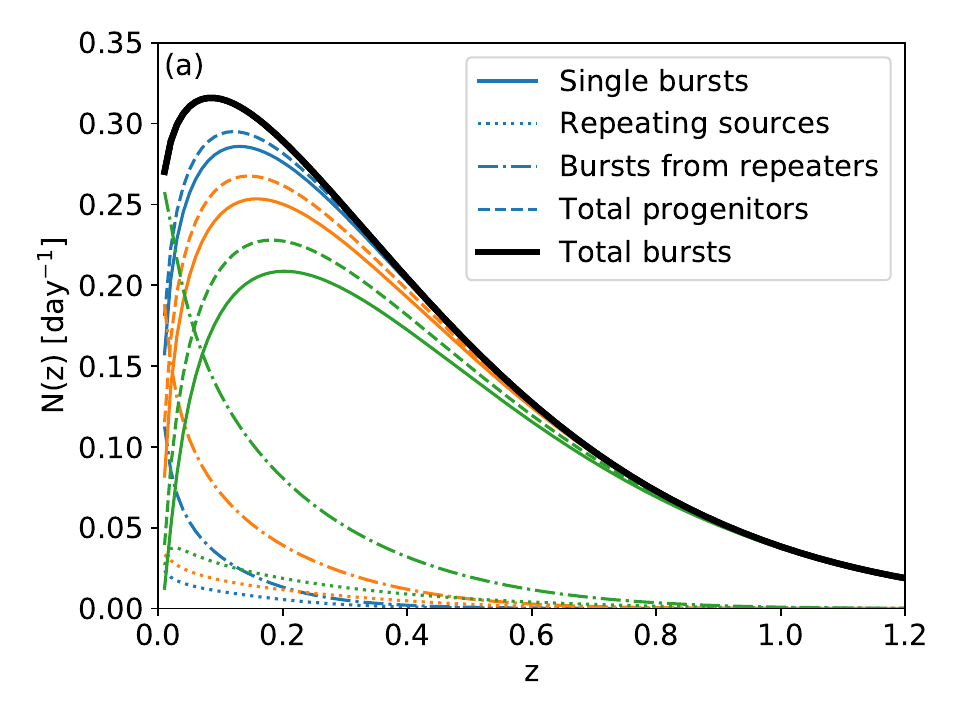}
    \includegraphics[width=\columnwidth]{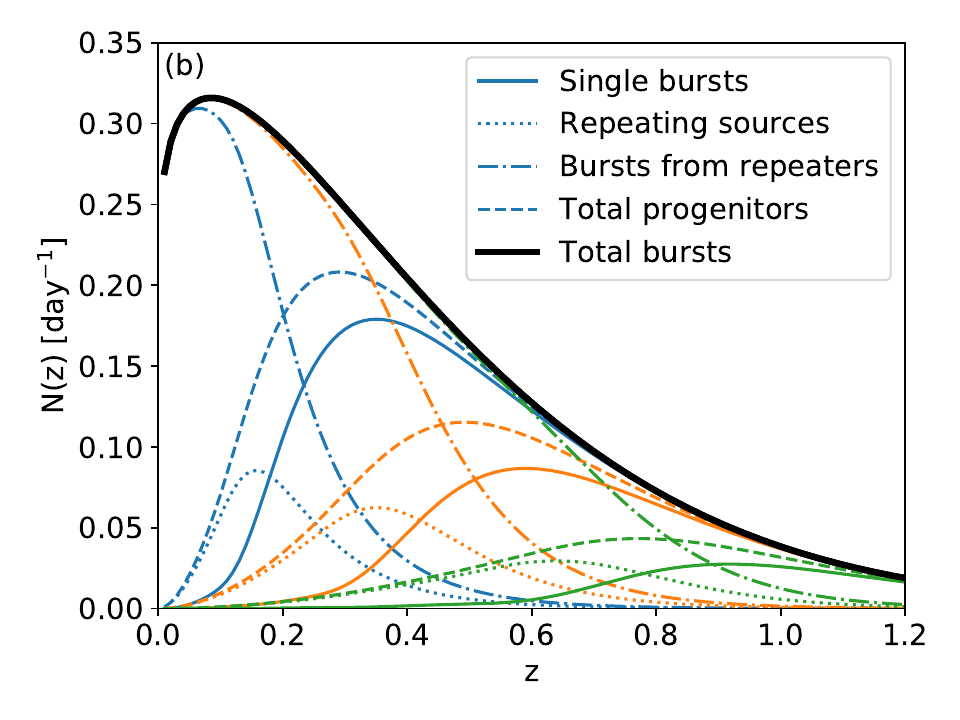}
    \caption{Effects of the repeating FRB population on the redshift distributon of FRBs expected from the CRAFT/ICS survey. Top: repeating FRBs with a broad distribution of rates; bottom: repeating FRBs identical to FRB~20121102A; observation times $T_f$, with lines appearing from left to right, are 10 days (red), 100 days (green), 1000 days (blue).}
    \label{fig:ics_example}
\end{figure}

The results for strong and distributed repeaters are shown in \figref{fig:ics_example}, in units of expected bursts per day. The total number of expected detected bursts is identical in all scenarios. However, all other measurable quantities depend on \tfield\ and the nature of the repeating FRB population. In all scenarios, intrinsically repeating FRBs (dotted lines) are more likely to be detected as such in the nearby Universe, which is the expected result. Necessarily, this decreases the expected number of once-off bursts in the local Universe (thin solid lines), since each FRB detected to repeat removes its first detected burst from the once-off distribution. The standard method of including repeating FRBs in population models, i.e.\ counting them only once using the first measured burst and adding apparently once-off bursts, is shown as ``Total progenitors'' (dashed lines). This method still results in a deficit of the modelled population in the local Universe compared to the total burst population. Only when including all bursts from measured repeaters (dot-dashed lines) as well as single bursts will the bias against low-redshift FRBs disappear.

The effect of \tfield\ is evident from Figure~\ref{fig:ics_example}. As \tfield\ increases (shown via changing colour), more intrinsically repeating FRBs in the nearby Universe are detected with multiple bursts. Thus, all distributions are pushed to higher $z$, with repeating FRBs taking an increasingly large fraction of the measured population. For the distributed repeaters scenario, this effect is small, and even after 1000\,days on a single field, by far the majority of FRBs are detected as once-off bursts, though the redshift peak for single bursts has shifted from 0.19 to 0.26 (with the unbiased peak for all bursts being at 0.15). For strong repeaters however, this effect is very important, with single bursts peaking at $z=$0.41, 0.67, and 0.95 respectively --- and even repeating sources peaking at higher redshifts than the true underlying total burst population. This latter effect is due to the small number of repeaters at low redshift being expected to produce a very large number of bursts, whereas for the distributed repeaters scenario, repetition remains dominated by rarely repeating objects at low redshift.

\begin{figure}
    \centering
    \includegraphics[width=\columnwidth]{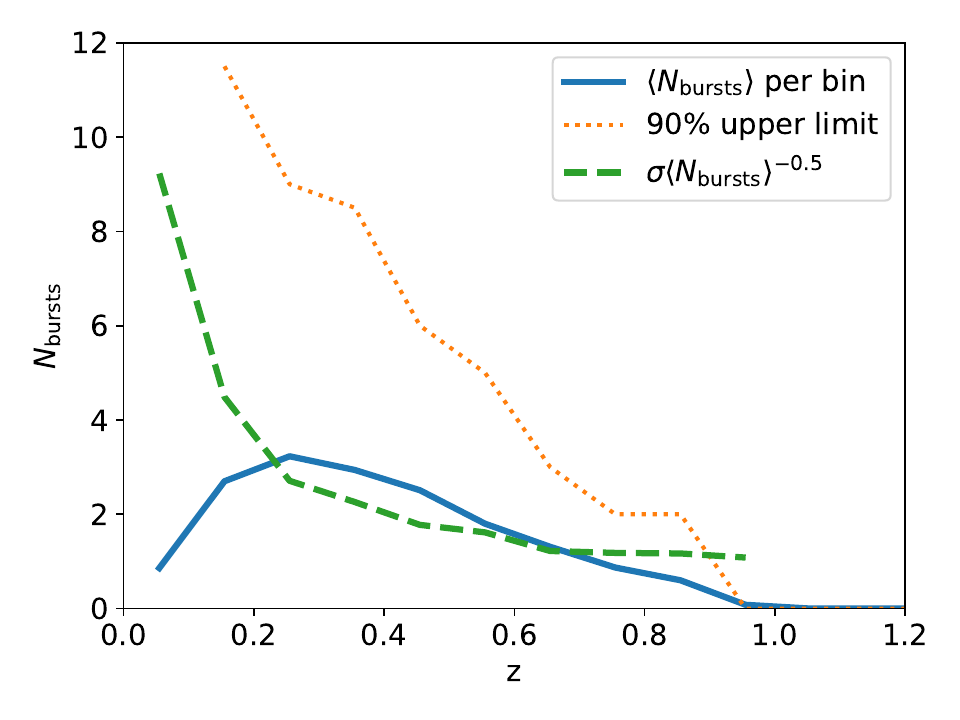}
    \caption{Burst statistics from 100 Monte Carlo simulations of an ASKAP/ICS 100 day pointing. Shown are the expected number of bursts $\left<N_{\rm bursts}\right>$, 90\% upper limit, and standard deviation normalised by the square root of $\left<N_{\rm bursts}\right>$, histogrammed as a function of redshift.}
    \label{fig:mc}
\end{figure}

Repeating FRBs also introduce significant cosmic variance into FRB surveys. Since the volume of the nearby Universe is small, there is a small chance for a strongly repeating FRB to be located in that volume --- however, if there is, many bursts will be detected from it due to its proximity. To illustrate this effect, a Monte Carlo sample of 100 instances of a 100-day ASKAP/ICS observation is generated, in a Universe consisting only of FRBs as strong as FRB~20121102A. The mean number of bursts detected, $\left< N_{\rm bursts} \right>$, is shown as a function of redshift in Figure~\ref{fig:mc}. As expected, it follows the shape of Figure~\ref{fig:ics_example}. For a Poisson distribution, the standard deviation $\sigma$ would be expected to scale with $\left< N_{\rm busts} \right>^{0.5}$ --- therefore, I normalise $\sigma$ by this value. At high redshift, it tends towards this expected value. However, this variance increases rapidly at low redshift. The number of bursts in the 90\% upper limit over all 100 simulations increases even more rapidly. The most extreme example of this variance is that in 99 of the 100 simulations, no bursts were simulated in the redshift interval $0 \le z \le 0.1$ --- however, in a single instance, 86 bursts from a single repeating FRB were simulated.

This presents a problem for population modelling. The requirement to include all bursts from repeating FRBs to avoid a bias will result in large stochastic fluctuations in population statistics, dominated by the small population of strongly repeating FRBs. The choice to do so, or not, represents a trade-off between bias and accuracy. I therefore proceed with the approach of modelling single bursts, and the number of repeaters, in this work, and do not directly model the number of bursts per repeater (though the distribution is fit via Monte Carlo in \secref{sec:fit_rates}). This results in zero bias, and only a small reduction in accuracy.

%% file: 04_modelling_chime.tex
\section{Modelling CHIME Catalogue~1 data}
\label{sec:CHIME}

The CHIME/FRB experiment is described by \citet{CHIME2018_system}. Not only does CHIME have the largest published sample of both repeating and non-repeating FRBs \citep{CHIME_catalog1_2021}, but since CHIME is not re-pointed to target specific sources, CHIME FRBs are detected in an unbiased manner. The downside however is that CHIME's angular resolution is relatively poor \cite[approximately 1'--10';][]{MichilliCHIMEbaseband}, so that the vast majority of FRBs are not localised with sufficient precision to identify their host galaxy. This makes CHIME data ideal for modelling the repeater vs non-repeater fraction, even if it is difficult to use it to model the FRB distribution in z-DM space. The CHIME experiment is modelled as below.

\subsection{Data sample}
\label{sec:data}

 Data is taken from \citet{CHIME_catalog1_2021} (hereafter, `\cat'), consisting of 536 FRBs. Nominally, this is divided into 474 once-off bursts, and 62 bursts from 18 repeaters.
 However, two repeaters --- FRB~20190417A, and FRB~20181119D from FRB~20121102A --- are only identified as such from observations external to \cat. Thus it is more proper to say that \cat\ identifies 16 repeaters and 476 non-repeaters, which is the statistic used in this work.
 
 CHIME have also recently published a search for repeating FRBs in an updated data-set, announcing the discovery of 25 new sources based on coincidences in DM--localisation space \citep{CHIME_2023_25reps}. I denote this the `\gold' sample, and discuss this in detail --- including why it is not used for fitting --- in \secref{sec:gold}.

 The analysis in \cat\ used three data quality cuts which are not implemented here, for the following reasons:

\begin{enumerate}
    \item Cut removing events with $\snr < 12$. This is implemented due to human inspection falsely rejecting low-\snr\ events. Since this work is primarily concerned with the ratio of repeaters to non-repeaters, rather than absolute FRB numbers, such a cut is not used. Repetitions were searched-for at a slightly lower (but undocumented) threshold than initial bursts \citep[][; see \secref{sec:gold} for further discussion of this effect]{CHIME_catalog1_2021}, and using the $\snr \ge 12$ cut would eliminate this effect. However, implementing it would leave only 7 of 16 repeating FRBs. I believe that the resulting loss of precision will be worse than the associated systematic error. 
    
    \item Cut removing events with $\DM < 1.5\,{\rm max}({\rm DM}_{\rm NE2001},$\\ $\DM_{\rm YMW16})$. Here, such events are retained --- it is true that \dmmw\ is poorly known, but this also means that the effect of such a cut is equally unknown.

    \item Far-sidelobe events. These will have poor localisation, and are identified using the FRB spectrum. However, this work is not concerned with localisation accuracy, but rather the number of FRBs detected, for which precise sidelobe details are not as important. Thus such a cut is not implemented.
\end{enumerate}

In the following analysis, the declination $\delta$ of each FRB, its DM, and whether or not it has been observed to repeat is used.

\subsection{Beamshape}
\label{sec:beamshape}

The CHIME beamshape is both unique and complex, with a broad primary beam with full-width at half-power of approximately 60$^\circ$--120$^\circ$ in the N--S direction and $1.3^{\circ}$--$2.5^{\circ}$ in the E--W direction \citep{CHIME2018_system,CHIME_catalog1_2021}. Within this envelope are 1024 coherently formed beams used for FRB detection, with full-width, half-max (FWHM) of 20'--40'. I use the frequency-dependent beamshape given in \citet{CHIME_catalog1_2021}, averaged over XX and YY polarisations, and implemented in the GitHub library {\sc chime-frb-beam-model}.\footnote{\url{https://github.com/chime-frb-open-data/chime-frb-beam-model}} This is sampled at 16 frequencies in a $300 \times 1000$ grid in RA, DEC, taking points within $8^{\circ}$ of the meridian. The final beamshape is taken as the envelope over all $1024$ formed beams after averaging over frequency. I discuss an alternative method in \ref{app:beamshape}.

In \citet{James2022Meth}, a telescope's beam pattern on sky is described via the `inverse beamshape', $\Omega(B)$, being the solid angle of sky viewed at any given sensitivity. The total time observing that solid angle is then a simple scaling constant between rate and the number of observes FRBs. When considering the response of a transit instrument such as CHIME to repeating FRBs, the relevant metric is $T(B)$, being the time spent observing a given sky position at beam sensitivity $B$ (relative to the nominal sensitivity at beam centre, where $B=1$). For a source at a given declination, $T(B)$ can be calculated from the RA-dependence of the beam pattern --- this is similar to the approach used by  \citet{Gardinier2021_frbpoppy_repeaters}.

\subsubsection{Declination dependence}
\label{sec:dec_dependence}

\begin{figure}
    \centering
    \includegraphics[width=\textwidth]{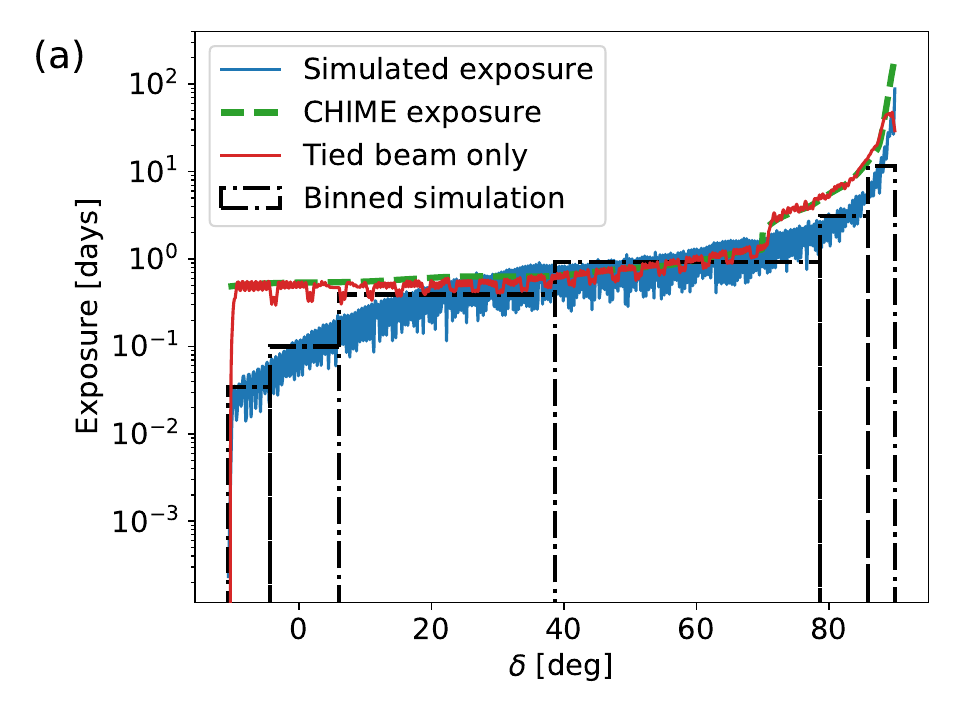}\\
    \includegraphics[width=\textwidth]{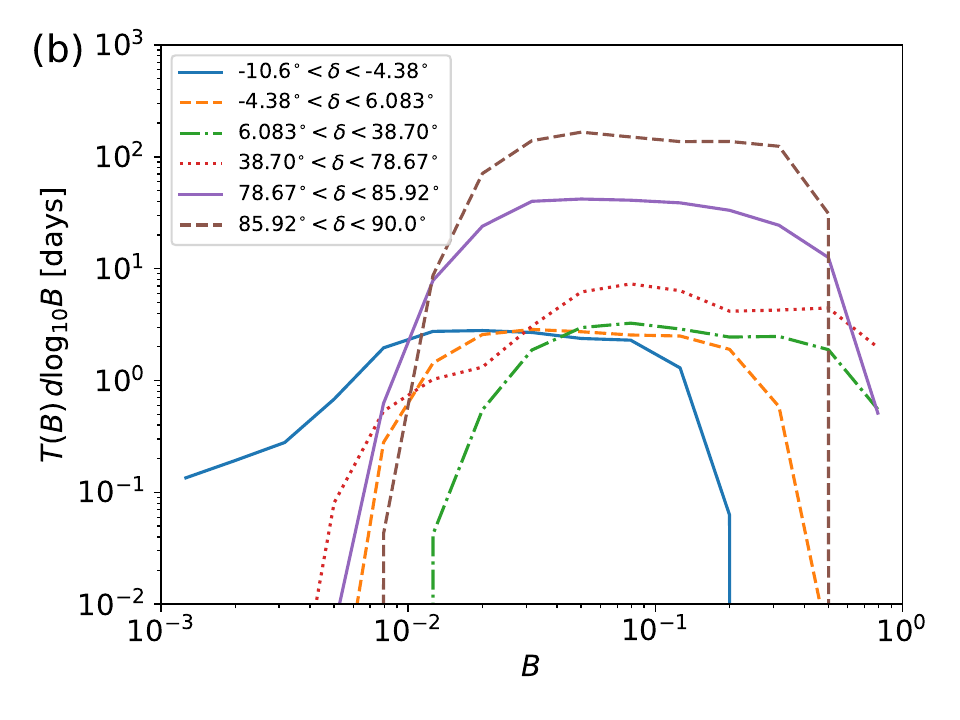}
    \caption{Top: declination-dependent exposure of the CHIME experiment --- simulation from this work based on the beamshape described in \citet{CHIME_catalog1_2021} and scaled using a total of 220 days' observation time, `CHIME' taken directly from \citet{CHIME_catalog1_2021}. Bottom: $T(\overline{B})$, calculated from the simulated beam pattern, and averaged over the indicated declination ranges.}
    \label{fig:exposure_comparison}
\end{figure}

I calculate the declination dependence of CHIME's beamshape by calculating $T(B)$ for 1000 declinations between $-11^{\circ}$ and $+90^{\circ}$. The effective exposure, $T_{\rm eff}$, is defined as the sum of times spent observing the source $T_i$, weighted by beam sensitivity $b$, assuming a Euclidean distribution of event rates:
\begin{eqnarray}
T_{\rm eff} & = & \sum_i T_i B^{1.5}. \label{eq:teff}
\end{eqnarray}
It is calculated for each declination as above, using discrete samples of source position (hence a sum, rather than an integral, in \eqref{eq:teff}), and is shown in \figref{fig:exposure_comparison}. The oscillatory behaviour is due to the 256 rows of tied beams, with sources passing over beam centre having significantly more exposure than those passing between beams. In the region $\delta \gtrsim 70^{\circ}$, the exposure is additionally increased by CHIME viewing sources transiting twice daily, both sides of the North Celestial Pole (NCP).
Also plotted is the CHIME exposure, taken from \citet{CHIME_catalog1_2021}, defined as the time in which a source is within the full-width half-max (FWHM) of a tied beam at 600\,MHz. This is used to normalise the total effective observation time to 311 days by using CHIME's simulation of the 600\,MHz tied beam, measuring the time a source spends within the FWHM, and fitting to the published CHIME exposure. This total effective time is less than the 342 day duration of the catalog, which is expected when allowing for equipment down-time.
Since CHIME's total exposure is a function only of the tied beamshape, it is therefore significantly greater than $T_{\rm eff}$ at low declinations where the primary beam is less sensitive.

For calculation purposes, declination is divided into six regions, spanning the full $[-11^{\circ},90^{\circ}]$ range. The bounds of each region are chosen so that the change in exposure between regions is not too large (at most a factor of three), while preserving reasonable statistics in each region. The mean exposures in each region are also shown as histograms in \figref{fig:exposure_comparison}.

\subsection{Sensitivity}
\label{sec:sensitivity}

The \zdm\ code by default uses a two-dimensional function of FRB width $w$ and DM to calculate the effective threshold $F_{\rm th}$ to an FRB with those properties. An important input into this is the telescope's time- and frequency-resolution used for FRB searches: in the case of CHIME, 0.983\,ms, and 0.0244\,MHz respectively \citep{CHIME_catalog1_2021}. Integrating over a modelled distribution of intrinsic FRB widths $w$ \citep{James2022Meth} and intrinsic scattering measures $\tau$ \citep{CHIME_catalog1_2021,James2022_H0}, this produces an efficiency function $\epsilon$ that acts to decrease measured \snr\, or equivalently, increase the effective detection threshold $F_{\rm th}$ above a nominal threshold $F_0$ in a DM-dependent manner. Including the effects of the beam $B$, $F_{\rm th}$ is given by
\begin{eqnarray}
F_{\rm th} & = & \frac{F_0}{B \epsilon(w,\tau,\DM)}. \label{eq:fth}
\end{eqnarray}

\begin{figure}
    \centering
    \includegraphics[width=0.9 \columnwidth]{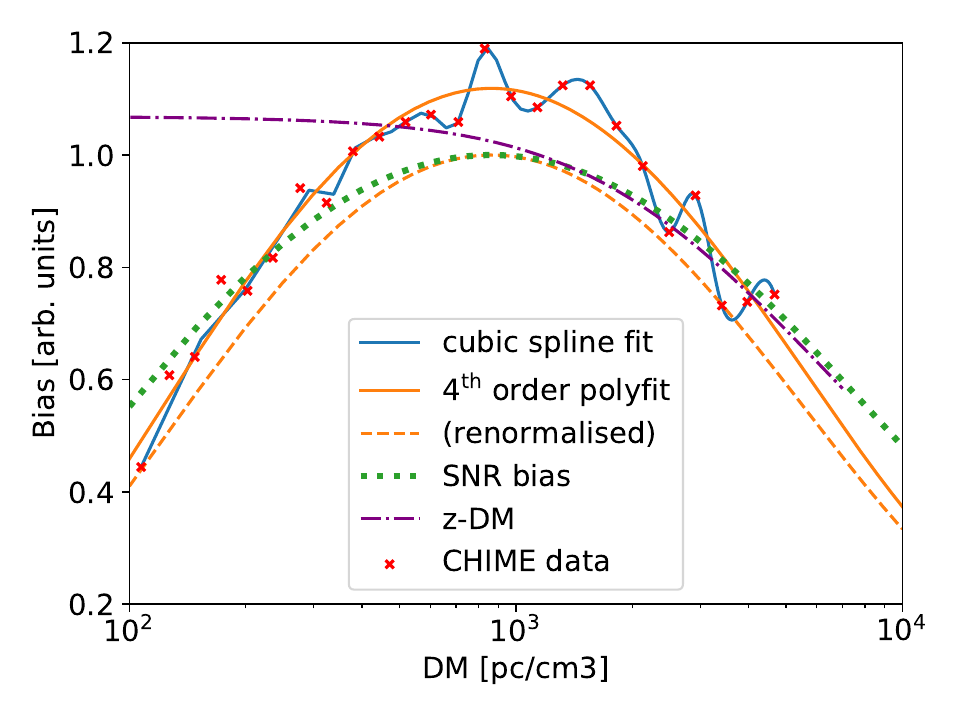}
    \caption{DM bias correction for CHIME data. Shown are values of $s$(DM) from \citet{CHIME_catalog1_2021} (red points), a cubic spline fit (blue solid line), the $4^{\rm th}$ order polynomial fit from this work (orange solid line), the renormalised fit (orange dashed line), and implied SNR bias (green dotted line).}
    \label{fig:chime_dm_bias}
\end{figure}

\citet{CHIME_catalog1_2021} extend this to a more complicated selection function, $P(\snr|F, DM, \tau, w, \gamma, r)$, giving \snr\ as a function of DM, $\tau$, $w$, $F$, and also including spectral shape parameters $\gamma$ and $r$. This is based on a sophisticated pulse injection system \citep{chime_injection_2022}, and thus accounts for not only the dedispersion code, {\sc bonsai}, but also the full detection pipeline, including RFI rejection.
The full selection function $P$ is not published, although it can presumably be inferred from the published library of injected pulses. Rather, one-dimensional selection functions, integrated over all other variables and modelled population probability distributions in those variables, are given. 

The DM selection function, $s(\DM)$, gives the relative fraction of FRBs passing selection cuts as a function of DM --- which is precisely what is required to model CHIME's DM distribution. To use CHIME's bias selection function, it is fit with a $4^{\rm th}$-order polynomial as shown in \figref{fig:chime_dm_bias}. It is then corrected to a peak value of unity, such that $N_{\rm obs} \le N_{\rm true}$, and converted to a modifier to the measured burst \snr\ by assuming a Euclidean relationship between event number and \snr, i.e.\ \snr$_{\rm bias} \sim s^{2/3}({\rm DM})$. This is compared to the efficiency function $\epsilon$ produced by the \zdm\ code, normalised such that the two efficiencies are equal at a DM of 1500\,\dmunits.

There are two clear regimes present in \figref{fig:chime_dm_bias}. Above 1000\,\dmunits, the CHIME \snr\ bias agrees well with that estimated by \zdm, which is indicative of sensitivity being fundamentally limited by the time--frequency resolution of the instrument. Below 1000\,\dmunits, \zdm\ flattens to represent efficient detection, while the efficiency of CHIME decreases. This is likely due to the effects of CHIME's system for mitigating RFI \citep{CHIME_catalog1_2021,chime_injection_2022}.

To estimate the fluence threshold at beam centre and peak DM efficiency, $F_0$, CHIME quotes a 95\% completeness threshold of 5\,Jy\,ms, compared to a theoretical minimum detection threshold of $\sim 1$\,Jy\,ms \citep{CHIME_catalog1_2021}. The factor of five between these thresholds is comparable to the factor of 6.6 by which the \zdm\ efficiency had to be increased to match the CHIME efficiency in \figref{fig:chime_dm_bias}. Thus the fitted \snr\ bias of CHIME's selection function has been implemented in the \zdm\ code, such that it acts in \eqref{eq:fth} as an efficiency factor; and $F_0=5$\,Jy\,ms to a 1\,ms burst is used. The value of the beam $B$ is parameterised according to \secref{sec:beamshape} and \secref{sec:dec_dependence}, so that repeating FRBs in each declination bin are modelled as being exposed to sensitivity thresholds $F_{\rm th}$ for times $T_{\rm obs}$.

%% file: 05_preliminary_results.tex
\section{Preliminary results --- initial comparison}
\label{sec:preliminary}

\input{Tables/extreme_params}

I begin by comparing the results of previous population modelling to CHIME single burst data. While the aim of the present 
manuscript is to model the relative once-off and repeat burst rates by varying the repeater properties of the population, the properties of the total FRB population --- luminosity function, source evolution etc.\ --- will be correlated. Furthermore, if a good fit for the single-burst population cannot be obtained, it will be impossible to determine whether goodness of fit for repeat bursts is a function of repeating or total population parameters.

FRB population parameters from \citet{Shin2022} and \citet{James2022_H0}, hereafter \shin\ and \jh\ respectively, are considered.
In the former case, only the best-fit set is used, since the allowed ranges are very broad. In the latter case, both the best fit, and sets compatible with the 90\% upper and lower limits of each parameter, are considered. To obtain these, the parameter in question is first set to its min/max value at 90\% confidence, and then a search is performed over all evaluated parameter sets with a similar parameter value for the best-fitting set of other parameters. These parameter sets are listed in \tabref{tab:extreme_params}.

\begin{figure}
    \centering
\includegraphics[width=\textwidth]{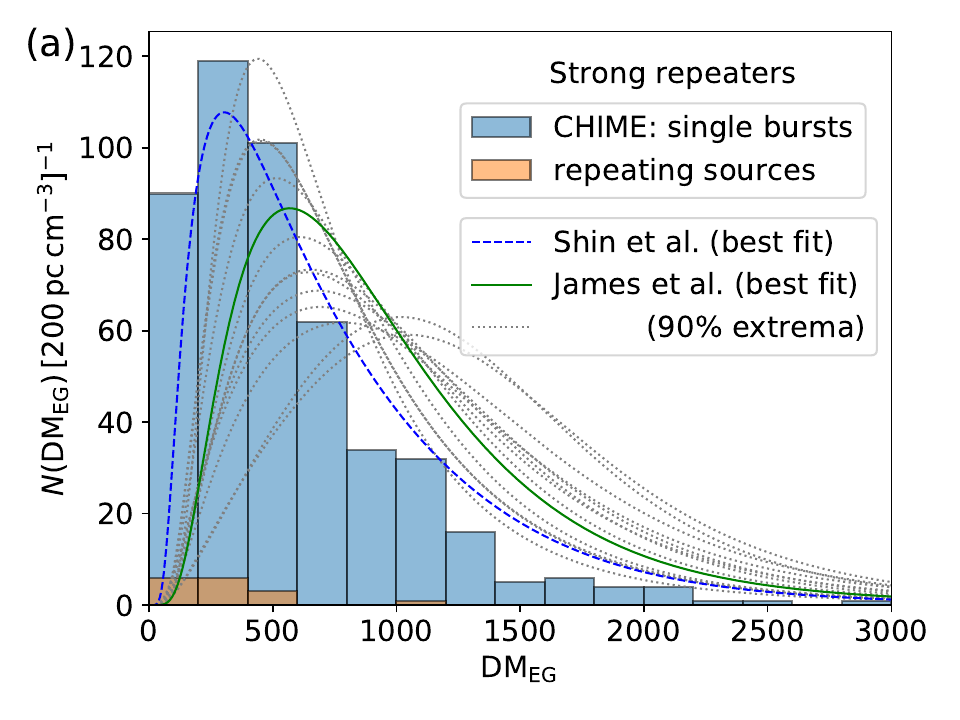}\\
\includegraphics[width=\textwidth]{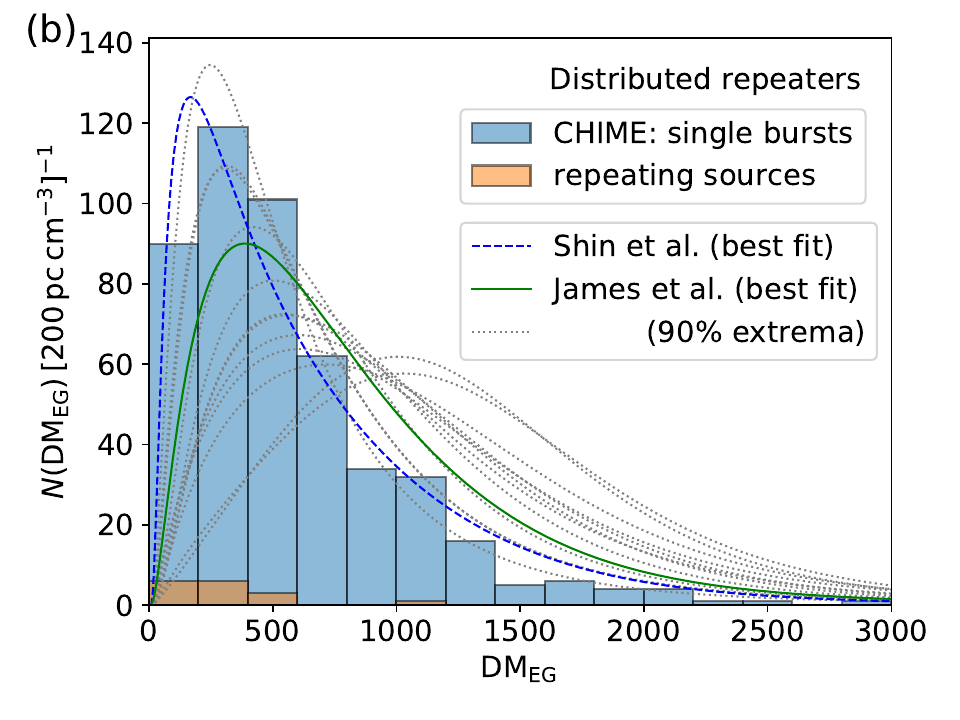}
    \caption{Observed rates of CHIME FRBs, showing sources observed as single and repeating, summed over declination. These are compared to estimates for the number of single bursts using the best-fit results from \citet{Shin2022} (solid, blue) and \citet{James2022_H0} (dashed, green), and 90\% extreme (dotted, grey) values of population parameters from Table~\ref{tab:extreme_params}, and assuming a population of repeating FRBs with `strong' (top) and `distributed' (bottom) repetition rates. Predicted singles rates are normalised to observed singles rates.}
    \label{fig:initial_chime_fit}
\end{figure}

Figure~\ref{fig:initial_chime_fit} shows fits to the single burst population for both the strong repeaters (top) and distributed repeaters (bottom) scenarios, for all considered population parameters. The most striking comparison is that models generally have difficulty fitting the large number of low-DM FRBs observed by CHIME, especially given the strong bias against low-DM events from the CHIME selection function. Performing 1-sample KS-tests \citep{kolmogorov,smirnov} of the CHIME data against the predicted curves using {\sc scipy}'s `stats' package, no parameter set provided a good fit to the strong repeater distribution. This is not surprising, since as discussed in \secref{sec:intro}, other analyses have already ruled out that all FRBs are strong repeaters. In the distributed repeaters scenario, only the minimum value of $\alpha = -1.91$ from \citet{James2022_H0} was compatible, with a p-value of $0.27$.

The fact that the parameters of \shin\ provide a reasonable fit (p-value of $0.0028$), but not the best fit, is a measure of the small but non-negligible systematic differences in the modelling. One possible cause is the use of $F_0$ set at the 95\% completion threshold, which is likely high. Reducing it would make CHIME more sensitive, and push the expected DM distribution to higher values, making it compatible with measurements. The model using the \shin\ parameter set also predicts a low number of singles bursts (131), though reducing the threshold to 2.5\,Jy\,ms to produce the correct number results in a poor fit to the DM distribution.

Another possibility is the complex interaction between burst shape and CHIME response, which is imperfectly captured here with the one-dimensional selection function $s({\rm DM})$, but which \shin\ model explicitly using injected pulses. A third possibility is that \shin\ fit to all progenitors (i.e.\ single and repeat FRBs). However, performing the same comparison here, the difference between the \shin\ predictions and data becomes greater.

Lastly, I also check that the fitting is not strongly affected by errors in \dmmw\ at low Galactic latitudes. Such an error would smear the \dmeg\ distribution, creating excess low-DM events. Re-doing the above analysis to include only FRBs with estimated Galactic contributions of less than 50~\dmunits\ changes the p-values by factors of order two, but this is minor compared to the different predictions between models. Thus this effect is ignored from hereon, and the entire sample is used to allow for greater precision and little cost of accuracy.

I therefore conclude that using the $\alpha$ `min' parameter set from \tabref{tab:extreme_params} in the \zdm\ code is likely to provide a reasonable fit to the CHIME single burst rate, and hence it is used to fit to the repeating FRB population in the following Section.

%% file: Tables/extreme_params.tex
\begin{table*}[hbt!]
\begin{threeparttable}
\caption{FRB population parameter sets used in this work. Shown are the best-fit parameter sets from \citet{Shin2022} and \citet{James2022_H0}, and a set of 12 parameter sets from \citet{James2022_H0} when each parameter takes the minimum/maximum value within its 90\% confidence interval.
The p-values from KS tests against the observed rate of CHIME single bursts is given as $p_{\rm KS}$ for different sets of FRB population parameters.}
\label{tab:extreme_params}
\begin{tabular}{c c c c c c c c c c c  }
\toprule
\headrow Scenario  & & \emax$^a$ & $\alpha^b$ & $\gamma^c$ & \sfrn$^d$ & $\log_{10} \muhost$${}^e$ & \sigmahost$^f$ & $\log_{10} C$$^g$ & $p_{\rm DM}^{\rm dist}$$^h$ & $p_{\rm DM}^{\rm strong}$$^i$ \\
                 &         & $\log_{10}$ [erg] & &        &        & \dmunits\ &       & ${\rm Gpc}^{-3}{\rm yr}^{-1}$ & & \\
\multicolumn{2}{l}{\citet{Shin2022}} & 41.38 & -1.39 & -1.3 & 0.96 & 1.93 & 0.41 & 4.99 & 0.0028 & $10^{-12}$ \\
\midrule
\multicolumn{2}{l}{\citet{James2022_H0}} & 41.26 & -1.0 & -0.95 & 1.13 & 2.27 & 0.55 & 4.47 & $10^{-22}$ & $10^{-54}$  \\
\midrule
\multirow{1}{*}{\emax}       & min & 41.0 & -1.0 & -0.7 & 1.0 & 2.3 & 0.6 & 4.63 & $10^{-21}$ & $10^{-38}$\\
       & max & 41.8 & -1.0 & -1.1 & 1.25 & 2.2 & 0.5 & 4.57 & $10^{-80}$ & $10^{-100}$  \\
       \midrule
\multirow{1}{*}{$\alpha$}       & min & 41.3 & -1.91 & -0.9 & 0.75 & 2.2 & 0.6 & 5.00 & $0.56$ & $10^{-23}$ \\
       & max & 41.3 & 0.24 & -0.9 & 0.75 & 2.2 & 0.6 & 4.94 & $10^{-7}$ & $10^{-26}$ \\
       \midrule
\multirow{1}{*}{$\gamma$}       & min & 41.8 & -1.5 & -1.18 & 1.75 & 2.2 & 0.5 & 4.35 & $10^{-125}$ & $10^{-138}$\\
       & max & 41.2 & -1.0 & -0.66 & 1.0 & 2.2 & 0.6 & 4.40 & $10^{-40}$ & $10^{-61}$  \\
       \midrule
 \multirow{1}{*}{\sfrn}      & min & 41.3 & 0.0 & -0.8 & 0.49 & 2.2 & 0.6 & 4.87 & $10^{-8}$ & $10^{-27}$  \\
       & max & 41.4 & -1.9 & -1.0 & 1.91 & 2.2 & 0.6 & 4.13 & $10^{-126}$ & $10^{-135}$ \\
       \midrule
\multirow{1}{*}{\muhost}       & min & 41.5 & -1.0 & -0.9 & 1.25 & 1.98 & 0.6 & 4.40 & $10^{-60}$ & $10^{-78}$\\
       & max & 41.3 & -1.0 & -0.9 & 1.0 & 2.45 & 0.6 & 4.69 & $10^{-53}$ & $10^{-74}$ \\
       \midrule
 \multirow{1}{*}{\sigmahost}      & min & 41.5 & -1.0 & -1.0 & 1.25 & 2.2 & 0.43 & 4.56 & $10^{-54}$ & $10^{-74}$\\
       & max & 41.4 & -1.0 & -0.9 & 1.25 & 2.2 & 0.82 & 4.50 & $10^{-65}$ & $10^{-83}$ \\
\bottomrule
\end{tabular}
\begin{tablenotes}[hang]
\item[a] Downturn ($\sim$`maximum') FRB energy, assuming a 1\,GHz rest-frame emission bandwidth.
\item[b] Frequency scaling of rate: $C \propto \nu^\alpha$.
\item[c] Cumulative power-law index of the FRB luminosity function.
\item[d] Scaling of FRB density with star-formation: $C \propto {\rm SFR}^{\sfrn}$.
\item[e] Log-mean of host galaxy DM contribution.
\item[f] Log-standard deviation of host galaxy DM contribution.
\item[g] Absolute total burst rate above $10^{39}$\,erg at 1.3\,GHz at $z=0$.
\item[h] p-value from the KS-test assuming the distributed repeaters scenario.
\item[i] p-value from the KS-test assuming the strong repeaters scenario.
\end{tablenotes}
\end{threeparttable}
\end{table*}

%% file: 06_fitting_results.tex
\section{Fitting results}
\label{sec:results}

To find a best-fit set of repeating FRB parameters (\rmin,\rmax,\rr), I first find the critical value of repetition rate, \rstar, such that when $\rmin = \rmax = \rstar$ (and thus the value of \rr\ is irrelevant), the correct number of repeating FRBs (in this case, 16) is reproduced. If $\rmin > \rstar$, then inevitably the model will produce too many repeating FRBs in the case that all bursts originate from repeaters; if $\rmax < \rstar$, then the model will not produce sufficiently many repeaters.

\begin{figure}
    \centering
    \includegraphics[width=\columnwidth]{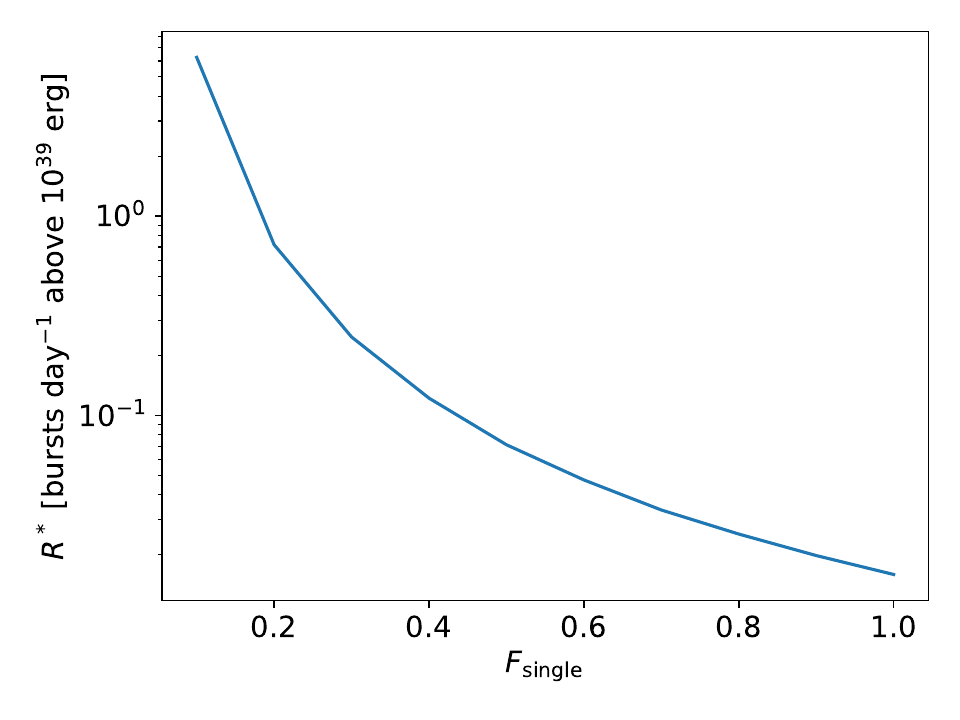}
    \caption{Critical value of repetition, \rstar, as a function of the fraction \FC\ of apparently once-off bursts that are attributed to repeaters.}
    \label{fig:rstar}
\end{figure}

To illustrate, in \figref{fig:rstar}, \rstar\ is plotted as a function of \FC, being the fraction of all apparently once-off CHIME FRBs that are produced by true repeaters. Reducing the number of once-off bursts attributed to repeaters, while keeping the observed number of repeaters in \cat\ constant at 16, means that a higher fraction of true repeaters get detected as such, i.e.\ they must be stronger. Indeed, for $\FC=0.1$, the average repeater must repeat about 6 times per day above $10^{39}$\,erg. For now, I continue with the case $\FC=1$ (i.e.\ all FRBs are repeaters), and revisit $\FC<1$ in \secref{sec:frep}.

\begin{figure}
    \centering
    \includegraphics[width=1.05\columnwidth,trim={0.5cm 0.2cm 0 0.8cm},clip]{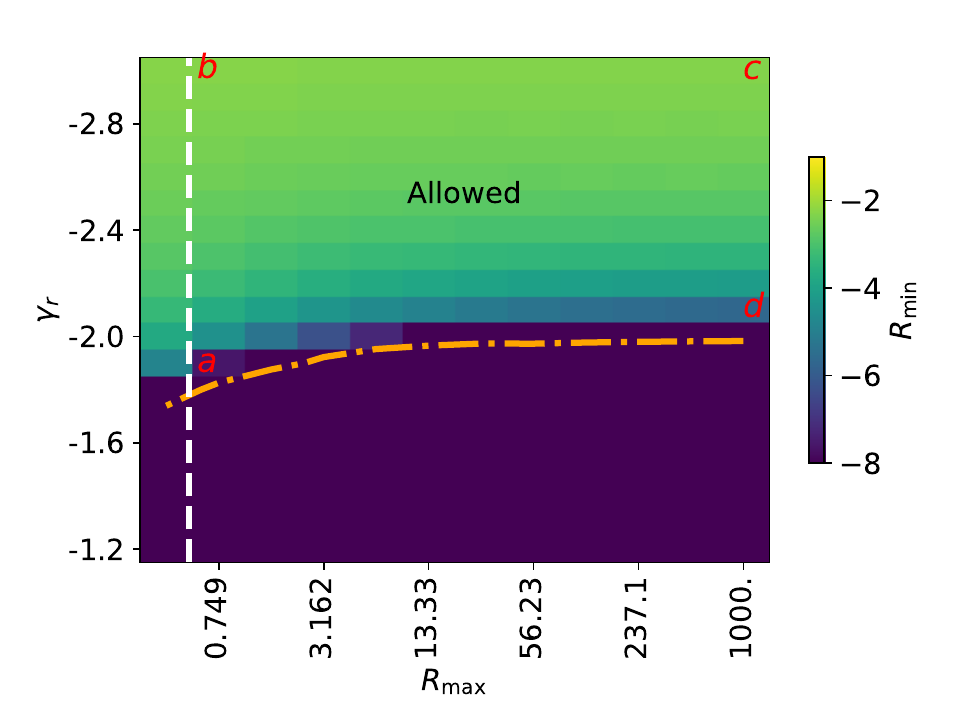}
    \caption{Value of \rmin\ producing the observed number of 17 repeating FRBs in the CHIME catalog \citep{CHIME_catalog1_2021} as a function of \rr\ and \rmax. Also shown are limits on \rmax\ (white dashed) from FRB 20180916B, and the region excluded as producing too many repeaters (orange dot-dash curve). The total `allowed' region is also indicated. Cases a--d used for \secref{sec:dmdist} and onwards are indicated in red.}
    \label{fig:rmin_example}
\end{figure}

For any two values in the set ($\rmin<\rstar$,$\rmax>\rstar$,\rr), and fixed \FC, the third value can be found such that the number of repeating FRBs observed by CHIME is reproduced exactly. For reasons to do with code optimisation, calculations varying \rr\ proceed very slowly, so \rr\ is held fixed to a small number of values. Since reasonable estimates of \rmax\ from observations of strong repeaters exist, it can be constrained to a sensible range. Therefore, I vary \rr\ and $\rmax > \rstar$, and for each, calculate the value \rmin. 

The resulting values of \rmin\ as a function of \rmax\ and \rr\ are shown in \figref{fig:rmin_example}. For steep \rr\, and for values of \rmax\ not much above \rstar, \rmin\ must also be close to \rstar (here, about 0.02 day$^{-1}$), while for flat \rr\ and large \rmax\ (the lower part of \figref{fig:rmin_example}), \rmin\ must be very small. For very flat \rr, the contribution of low-R repeaters to the apparently once-off burst rate becomes sufficiently negligible that it cannot `dilute' the number repeaters observed as such when \rmax\ is large. This excludes the region in the lower right of the figure. For calculation purposes, $\rmin$ is set to $10^{-8}$ in this region --- even though this over-produces the number of repeaters, it allows calculations to compare their DM, declination, and repeat rate distributions.

The lower limit on \rmax\ given by FRB~20180916B is also shown --- at a distance of approximately 150\,Mpc \citep{MarcoteRepeaterLocalisation2020}, its observed repetition rate above 5\,Jy\,ms of $0.448^{+0.1}_{-0.086}$\,hr$^{-1}$ \citep{CHIME_2023_25reps} approximately translates to a rate above $10^{30}$\,\eh\ (i.e., $10^{39}$ erg assuming a 1\,GHz bandwidth) of 0.5\,day$^{-1}$ when using a cumulative fluence index $\gamma=-1.5$.

An upper limit on \rmin\ can be estimated from the lowest estimated rate for a low-DM FRB observed by CHIME. FRB~20190518D \citep{CHIME_catalog1_2021} has a DM of $202.2$\,\pccc, with an estimated contribution by the Milky Way's interstellar medium (ISM) of 53.7\,\pccc\ according to the NE2001 model \citep{CordesLazio01}. Conservatively assuming a low combined halo and host DM contribution of 35\,\pccc, similar to that found for FRB~20200120E \citep{CHIME_M81_2021}, and using a simplistic model of $z \sim 10^{-3} \, \dmeg$, produces an approximate maximum redshift of $z \sim 0.11$. Again using $\gamma=-1.5$, this produces an upper limit on \rmin\ of $0.05$\,day$^{-1}$.

The combination of these three constraints produces the allowed region shown in \figref{fig:rmin_example}. Note that in all cases, \rmin\ is below (i.e.\ compatible with) the limit from FRB~20200120E.

\subsection{Dispersion measure distribution}
\label{sec:dmdist}

\begin{figure}
    \centering
    \includegraphics[width=\columnwidth]{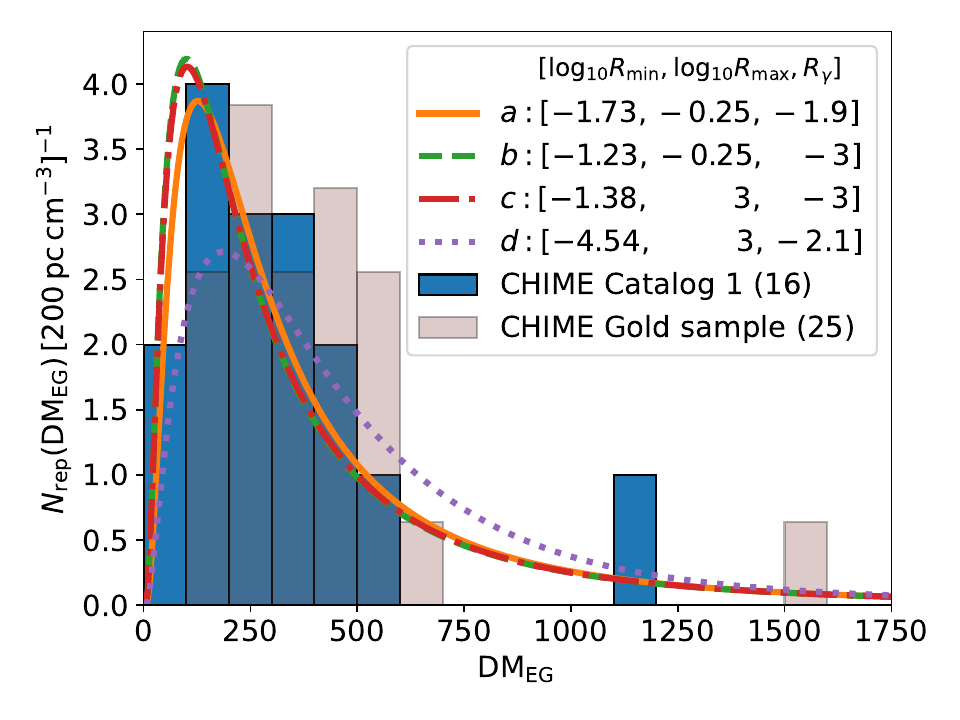}
    \caption{Predicted DM distribution of repeating FRBs compared to that from CHIME catalog 1 \citep{CHIME_catalog1_2021}, calculated using cases $a$--$d$ from \figref{fig:rmin_example}, and the golden sample of repeaters from \citet{CHIME_2023_25reps} (renormalised to 16). Note that $b$ and $c$ overlap.}
    \label{fig:dm_comparison}
\end{figure}

Within the range allowed by \figref{fig:rmin_example}, the z--DM distribution predicted for each will in-general be different. To illustrate, I take four scenarios, $a$--$d$, from the corners of the allowed region. In each case, the predicted DM distribution of repeating FRBs is plotted in \figref{fig:dm_comparison}.

From \figref{fig:dm_comparison}, it can immediately be seen that strong repeaters are more likely to be found at large distances (higher DM$_{\rm EG}$). Case $d$ has a significantly higher distribution of DMs compared to the other three, and it is the only case with a significant number of strong repeaters (cases $a$ and $b$ have low \rmax, while case $c$ has such a steep \rr\ that the number of strong repeaters is negligible).

I also compare these DM distributions with those found from the \cat\ and \gold\ samples. While formally the DM distribution of the \gold\ sample is statistically consistent with that of \cat\ \citep{CHIME_2023_25reps}, the distribution is biased (see \secref{sec:gold}). That case $d$ well-reproduces the \gold\ sample DM distribution therefore should not be taken as evidence for it.

This comparison highlights a prediction of all repeating FRB models, which is the stronger upward skew of the DM distribution compared to single FRBs. This effect is seen in CHIME data, with two (one) high-DM repeaters in the \cat\ (\gold) samples.

\begin{figure}
    \centering
    \includegraphics[width=1.05\columnwidth,trim={0.5cm 0.2cm 0 0.8cm},clip]{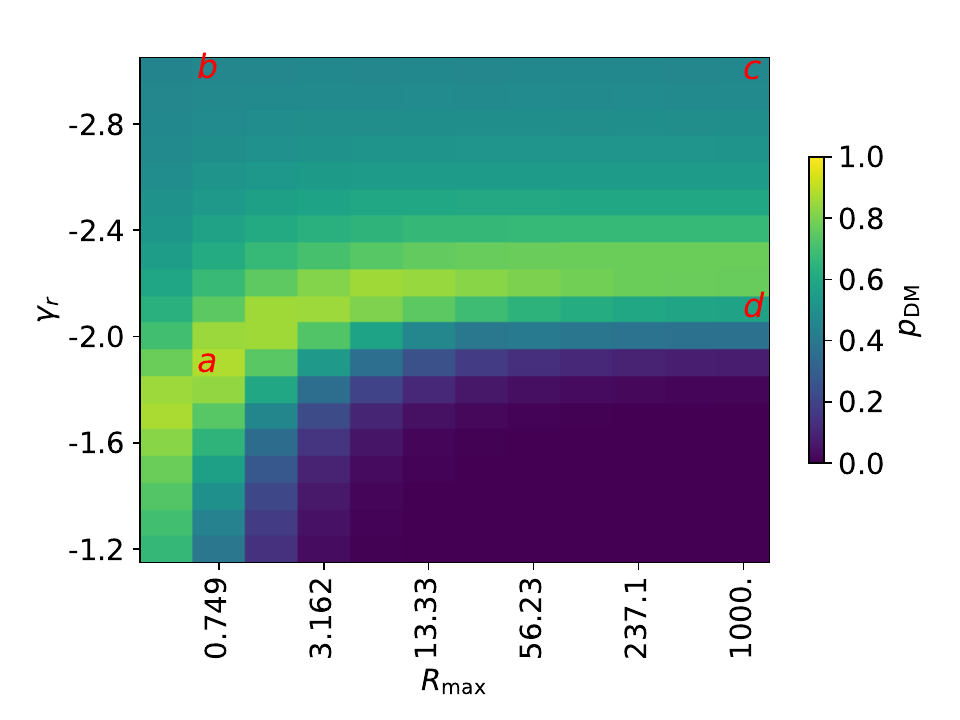}
    \caption{P-values from a KS test of the DM distribution of \cat\ repeating FRBs, $p_{\rm ks}({\rm DM}_r)$, against predictions from models with different values of \rmax\ and \rr. Other features are identical to  \figref{fig:rmin_example}, including cases $b$ and $c$ overlapping.}
    \label{fig:dm_ks}
\end{figure}

To quantify agreement in DM space, a KS-test using the \cat\ FRBs and predicted DM distributions over \rmax, \rr\ space is performed. The results are shown in \figref{fig:dm_ks}.

An implicit assumption of the above analysis is that the intrinsic distribution of \dmhost\ is identical regardless of the repetition rate of repeaters. The observation of persistent radio sources at the locations of at least two bright repeaters \citep{Marcote2017PRS,Niu2022} suggests that these presumably young objects would be more likely to have a larger \dmhost, bearing in mind that this term includes material in the vicinity of the progenitor, as well as the host galaxy's ISM and halo contributions. This could then be responsible for observing repeating FRBs (which are on-average intrinsically stronger repeaters) to have slightly more DM than expected, and would not constitute hard evidence against the model. This might be the case for the $\rr \lesssim -2.4$ region of \figref{fig:dm_ks}, which is slightly disfavoured because it over-predicts the number of low-DM repeating FRBs. However, should observed repeaters have less DM than expected, this is clear evidence to reject the model. This is the case for the already ruled-out lower region of \figref{fig:dm_ks}, which predicts more high-DM repeating FRBs than observed.

\subsection{Declination distribution}

The declination distribution of CHIME FRBs also holds information that allows us to discriminate between scenarios. For repeating FRBs, the difference between observing a small patch of the sky around the North Celestial Pole almost continuously, and surveying several steradians near the equator for only a few minutes each day, is very important. Near the equator, only the strongest repeaters will be detected as such, while near the Pole, the small probed volume makes observations subject to cosmic variance.

\begin{figure}
    \centering
    \includegraphics[width=\columnwidth]{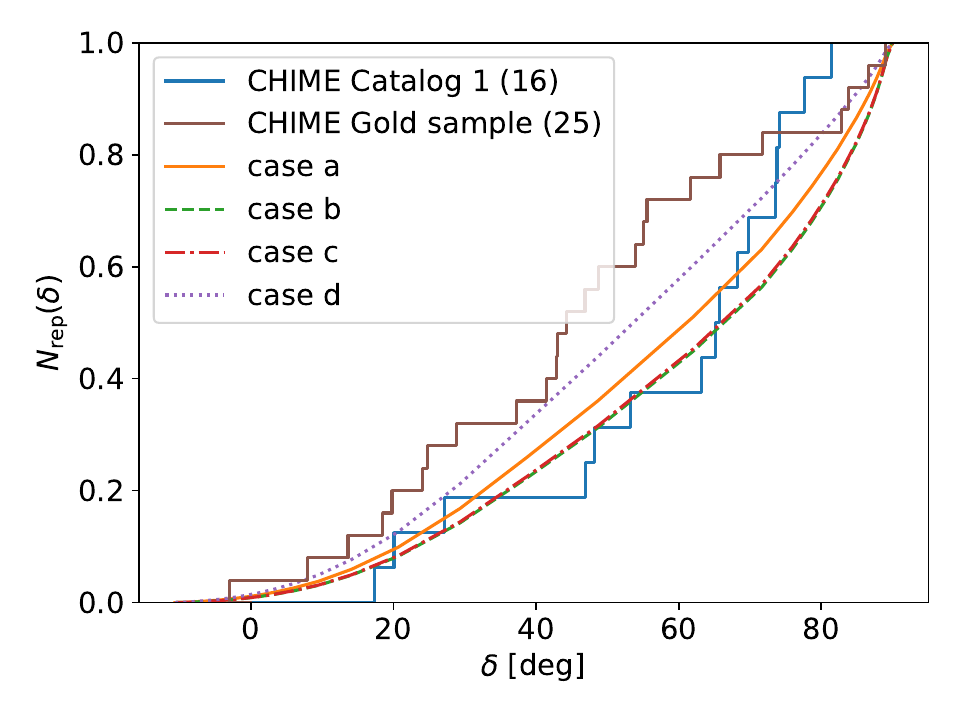}
    \caption{Cumulative histogram of the CHIME repeating FRB declination ($\delta$) distribution, for both the \cat\ and \gold\ samples, compared to Monte Carlo predictions from four example cases.}
    \label{fig:dec_distribution}
\end{figure}

In Figure~\ref{fig:dec_distribution}, I plot the declination distribution of CHIME once-off and repeating FRBs, and compare this against model predictions. For this plot, the number of declination bins into which CHIME was divided was increased to 30, whereas six declination bins was found to be sufficient to model the total number of repeaters, and their DM distribution.

Since the x-axis of \figref{fig:dec_distribution} is increasing linearly with $\delta$, most of the solid angle is concentrated on the left-hand-side of the figure. Despite this, the number of repeaters --- both observed and predicted --- increases as fast as, or faster, than linearly with $\delta$. That the \gold\ sample shows the least steep rise with $\delta$ is likely because of the previously discussed bias against high declinations due to the increased background rate. This is evidence that the repeating population is dominated by progenitors with low apparent repetition rates that are best probed with deep observations (i.e.\ at high declinations), rather than sources with high apparent rates that are best detected in broad shallow surveys (i.e.\ at low declinations).

Of the four cases analysed, a--c show good agreement with \cat\ in the $\delta \lesssim 60^{\circ}$ range, while not even d can match the rapid rise in repeater rates above this range. This suggests a simple fluctuation in the data, either a deficit at low declinations, or an excess at high declinatons --- though an alternative explanation is the influence of non-Poissonian repetition (see \secref{sec:nonpoissonian}).

\begin{figure}
    \centering
    \includegraphics[width=1.05\columnwidth,trim={0.5cm 0.2cm 0 0.8cm},clip]{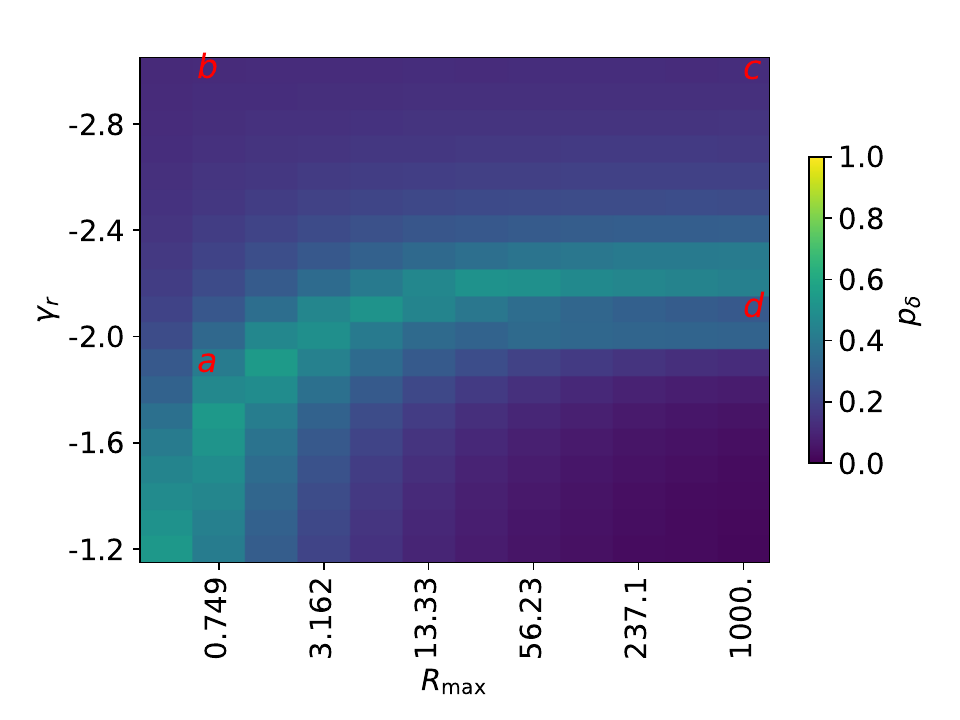}
    \caption{Results of the KS-test against the declination distribution of identified repeating FRBs. Shown is the p-value as a function of \rmax\ and \rr.}
    \label{fig:ksdec}
\end{figure}

I characterise the agreement in declination distributions via a KS-test, with associated p-values given in \figref{fig:ksdec}. The greatest discrepancy with data is the aforementioned excess of high-$\delta$ repeaters, and the upper region of the figure is disfavoured because it reproduces this particularly poorly. The lower right region is disfavoured because this predicts mostly bright repeaters that should be found in the greater region of sky viewed at low declinations.

\subsection{Repetition rate distribution}
\label{sec:fit_rates}

Most repeating CHIME FRBs are not localised, so that scaling between intrinsic and apparent repetition rates, which requires the luminosity distance to be known, is not possible. This precludes a direct fit to the rate distribution. Nonetheless, different combinations of \rmin, \rmax, and \rr\ lead to more/less repeating FRBs being observed with different apparent repetition rates.

Directly computing the number of FRBs with any given repetition rate is highly inefficient however --- the algorithm currently estimates the number of repeaters by explicitly calculating $N_0$ and $N_1$ only. Extending this to a large number of $N_{\rm burst}$ values is computationally prohibitive. A Monte Carlo sampling algorithm was therefore implemented that generates repeating FRBs according to their underlying modelled distribution in z--DM--$R$ space, and simulates the number of observed bursts assuming a Poissonian distribution.

\begin{figure}
    \centering
    \includegraphics[width=\columnwidth]{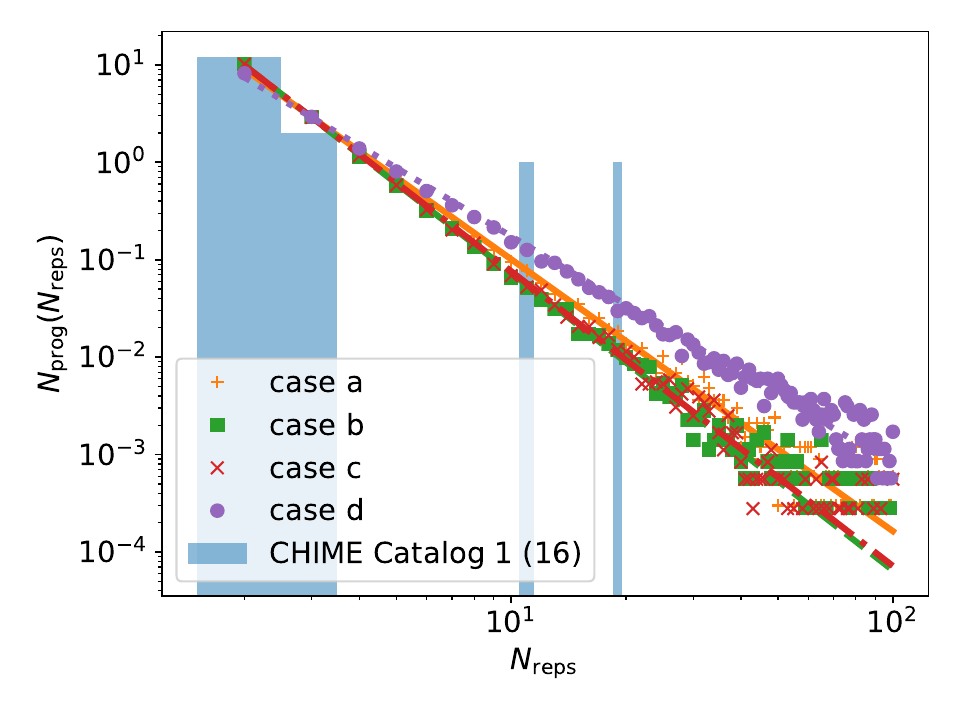}
      \includegraphics[width=\columnwidth]{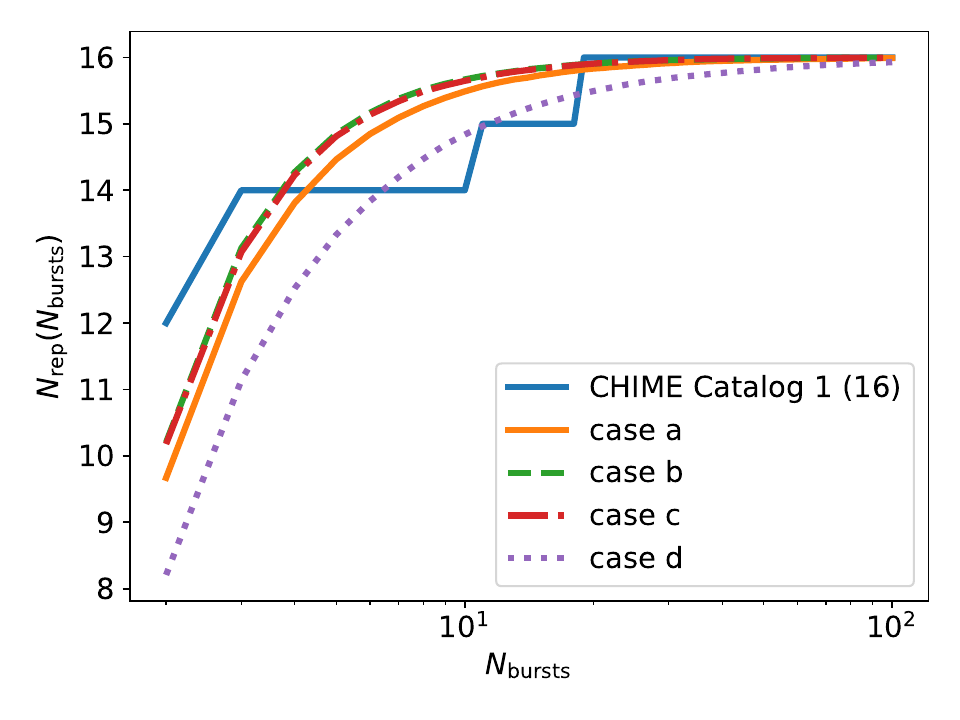}
    \caption{Top: histogram of observed number of repetitions in CHIME repeating FRBs from \cat, compared to Monte Carlo predictions from four example cases, a--d (points). A power-law fit (lines) is given for each. Bottom: the same data, but shown as a cumulative distribution.}
    \label{fig:mc_histogram}
\end{figure}

\begin{figure}
    \centering
    \includegraphics[width=1.05\columnwidth,trim={0.5cm 0.2cm 0 0.8cm},clip]{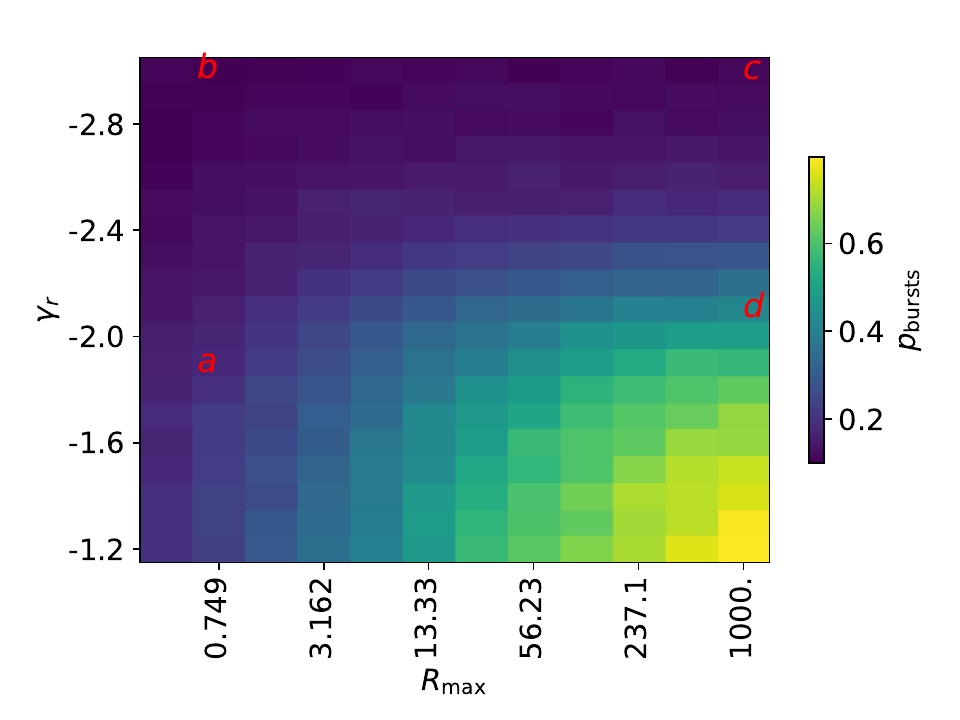}
    \caption{Maximum-likelihood estimates of FRB repeat parameters based on the distribution of the number of observed bursts from each repeater in the \cat\ sample.}
    \label{fig:mc_pvalues}
\end{figure}

To overcome Monte Carlo fluctuations, at least 1000 times as many repeating FRBs as expected are simulated, and a histogram produced in terms of the observed number of bursts by CHIME. This is then fit with a power-law distribution, and for histogram bins with less than 10 simulated repeaters, the observed number is replaced with the fitted number for purposes of evaluating likelihoods. An example of this procedure is shown in \figref{fig:mc_histogram}.

Since the data are discrete (integer numbers of bursts only), a KS-test
to assign a goodness-of-fit is inapplicable. Instead, the likelihood of the observed histogram of $N_{\rm burst}$ values for the 16 CHIME repeating FRBs from the \cat\ sample is calculated, given predictions from the Monte Carlo histogram. This is then repeated for at least 1000 sets of 16 Monte Carlo FRBs, and the fraction of likelihoods that are lower than that observed is determined. This produces a p-value, $p_{\rm bursts}$, under the null hypothesis that the Monte Carlo sample is the truth. Results are plotted in \figref{fig:mc_pvalues}.

The repeat-rate distribution is best-reproduced by models with a large number of bright FRBs, since the two CHIME FRBs with high repetition rates in \cat\ --- FRB20180814A (11 bursts), and FRB20180916B (19 bursts) --- are difficult to reproduce with models of low \rmax\ and/or steep \rr.

\subsection{Combined likelihood}
\label{sec:ptot}

\begin{figure}
    \centering
    \includegraphics[width=1.05\columnwidth,trim={0.5cm 0 0 0},clip]{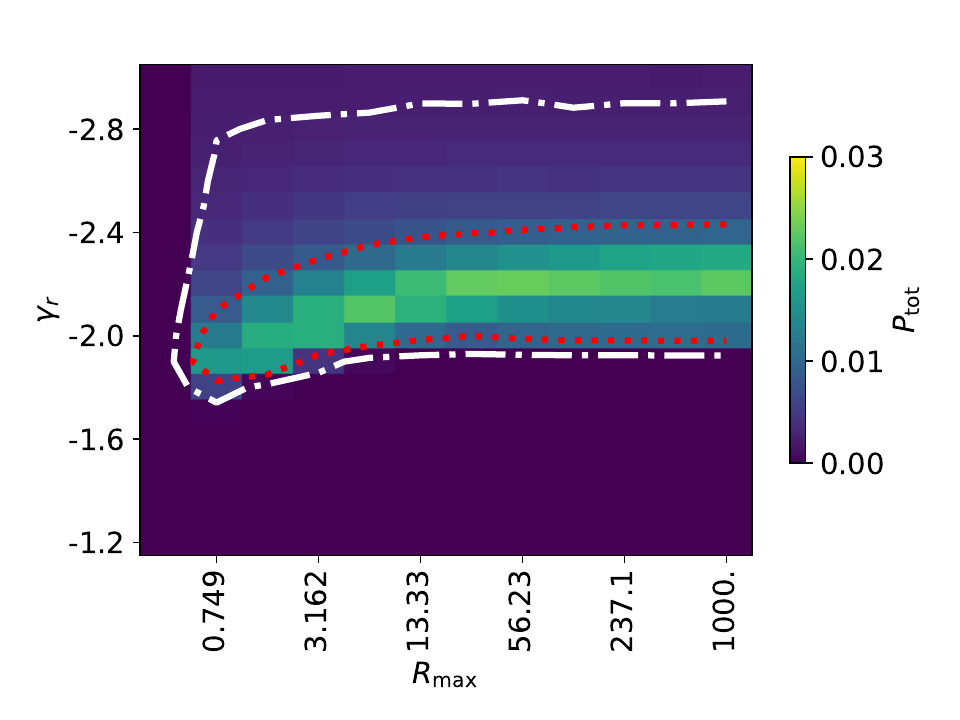}
    \caption{Posterior probability of repeating FRB parameters assuming that all FRBs repeat. Shown are 68\% (red dotted lines) and 95\% (white dot-dash lines) confidence intervals.}
    \label{fig:ptot}
\end{figure}

Combining the evidence from the DM, $\delta$, and $N_{\rm burst}$ probabilities derived above, the combined probability $p_{\rm tot}$ is constructed as
\begin{eqnarray}
p_{\rm tot} & = & p_N \, p_\delta \, p_{\rm DM} \, p_{\rm bursts},
\end{eqnarray}
where $p_N$ is a Poissonian probability of observing 16 repeaters, which suppresses the region of the parameter space that over-produces repeaters.
The probabilities are renormalised to sum to unity over the investigated range, excluding $\rmax < 0.5$\,day$^{-1}$, and confidence intervals assuming flat priors in \rr\ and $\log \rmax$ are constructed.
This results in the probability distribution, and confidence intervals (C.I.s), shown in \figref{fig:ptot}.

The 95\% C.I.\ encompasses almost the entire allowed region from \figref{fig:rmin_example}, showing that DM, $\delta$, and $N_{\rm burst}$ are not strong discriminators between different models of the repeating FRB population. However, a preference for $\rr=-2.2_{-0.8}^{+0.6}$ (68\% C.I.), and $\rmax \ge 0.75$, is found.

The above formulation for $p_{\rm tot}$ ignores correlations between variables: closer (low DM) FRBs, and those viewed closer to the zenith, will be more likely to have more bursts detected. A better analysis would use the full 3D distribution of $p({\rm DM}, \delta, N_{\rm bursts})$, similarly to the use of redshift, DM, and burst energy in standard z--DM analyses. However, generating the distribution of $N_{\rm bursts}$ is computationally intensive, and thus it is only performed in one dimension. This method should be revisited once other systematic effects, as discussed in \secref{sec:systematics}, are treated.

\subsection{What if not all FRBs are repeaters?}
\label{sec:frep}

It is of course possible that repeating FRBs do not constitute the total FRB population. Evidence for this comes from the different spectro-temporal properties of repeaters compared to non-repeaters \citep{CHIME_morphology_2021}, and a tentative association of FRB 20190425A with binary neutron star merger GW190425 \citep{Moroianu2023}. If such a population exists, it is likely subdominant --- most cataclysmic events, which would produce intrinsically once-off FRBs, have a rate which is much too low to explain the total FRB rate \citep{Ravi2019Prevalence}. Therefore, observations of the total FRB population still serve as good constraints on the total repeating population, and the predictions made here remain valid. Nonetheless, in this section, the case where repeating FRBs are responsible for a sub-dominant fraction of the total number of bursts observed by CHIME is investigated.

\begin{figure}
    \centering\includegraphics[width=\columnwidth]{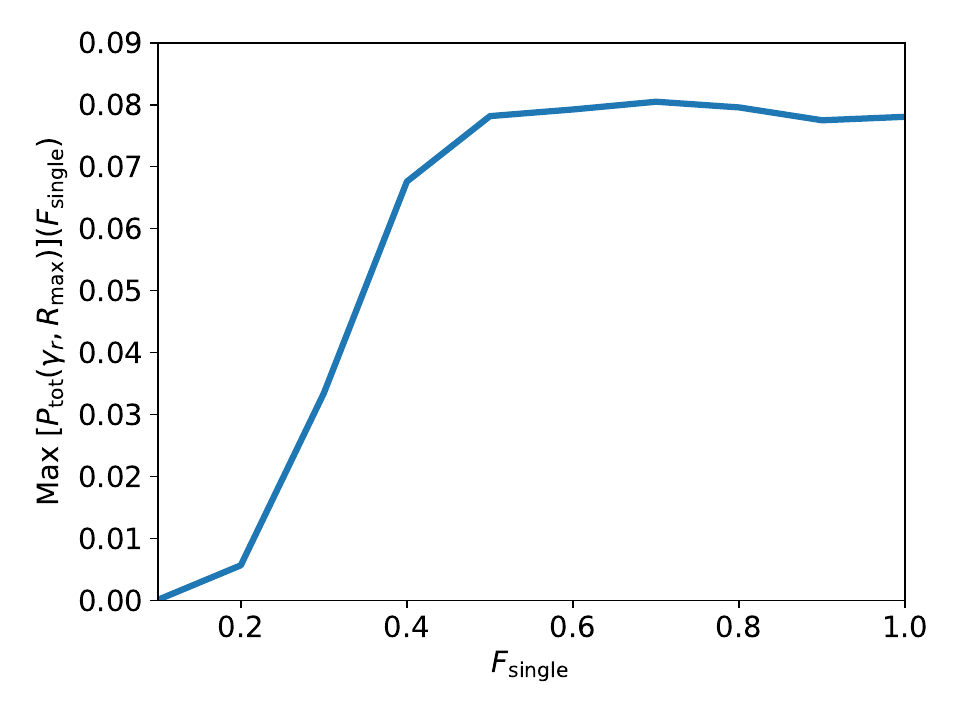}
    \caption{Maximum value of the joint probability $P_{\rm tot}$ over all analysed \rr, \rmax, as a function of the fraction of all CHIME single bursts explained by repeating FRBs, \FC.}
    \label{fig:Fptot}
\end{figure}

The above analysis is repeated by first optimising \rmin\ to produce 16 CHIME repeaters and some fraction \FC\ of the total singles burst rate, and calculating the joint probability $P_{\rm tot}(\rr, \rmax, \FC)$. The peak likelihood over \rr\ and \rmax\ for each \FC, Max$[P_{\rm tot}(\rr, \rmax)](\FC)$, is then plotted in \figref{fig:Fptot}.

In the range $0.5 \le \FC \le 1$, the peak probability is essentially identical, with fluctuations likely due to the coarse gridding in \rr--\rmax\ space. The likelihood decreases for lower values of \FC\ --- this is driven almost entirely by $p_{\rm DM}$, since decreasing \FC\ increases the fraction of true repeaters detected as such, which requires on-average stronger repeaters that are invariably detectable at greater distances. This pushes the predicted DM distribution to higher values, inconsistent with CHIME data. 

A note of caution is warranted however: if a small fraction of all single bursts are produced by repeaters, then the assumption that repeating FRB population parameters are the same as that of the total population is a bad one. Therefore, while it can be concluded that these results are consistent with a best-fit of all FRBs being from repeaters, it cannot be concluded that this excludes a large fraction of FRBs being from intrinsically once-off events.

%% file: 07_future_prospects.tex
\section{Future prospects}
\label{sec:future_prospects}

Now that an estimate of the parameters of the repeating FRB population has been made, I make predictions for the effects of repetition on future observations. In the following, cases $d$ (close to the best-fit values found in \secref{sec:results}) and $b$ (marginally excluded at the 90\% level, albeit when considering random error only) are considered as two significantly different, but plausible, cases.

\subsection{Rate of new repeater discoveries with CHIME}
\label{sec:tdist}

\begin{figure}
    \centering
    \includegraphics[width=\columnwidth]{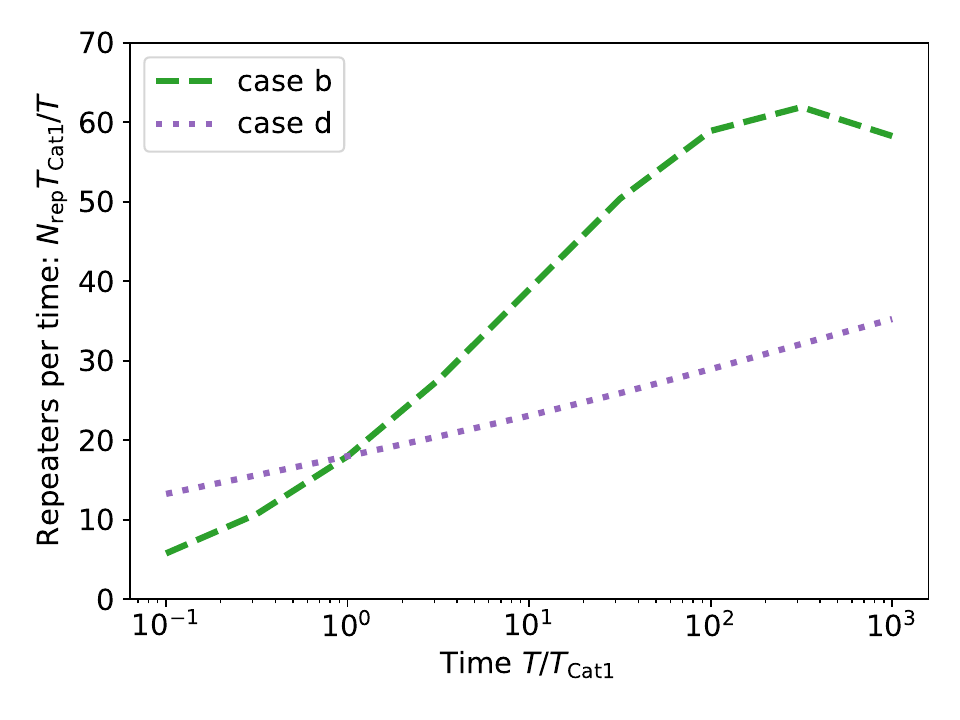}
    \caption{Number of repeaters, $N_{\rm rep}$, normalised by total observation time $T$, in units (and as a function) of the exposure from \cat, for cases $b$ and $d$.}
    \label{fig:time_effect}
\end{figure}

As time spent observing a particular field increases, the number of repeating FRBs should eventually saturate, as essentially all such objects in the field are detected. Seeing the rate of detected repeaters plateau at a level where a large number of once-off bursts have no associated repeater would be a clear indication of two populations. This raises the question: how long might CHIME have to wait until the rate of new repeating FRB detections decreases?

The answer is a very long time. Regardless of the scenario under consideration, the number of repeating FRB progenitors at high redshifts will vastly outnumber those at low redshifts due to the increased volume of the Universe. As observation time increases, the number of repeating FRBs in the nearby Universe will saturate, but the rate of repeater discoveries --- both as single and repeat bursts --- in the distant Universe increases. This effect is seen in \figref{fig:ics_example} --- in \figref{fig:time_effect}, this is simulated for CHIME, by simply increasing the observation time in units of \cat, $T_{\cat}$ (which is approximately a year's worth of exposure). In case $b$, there are relatively few strong repeaters, and saturation is expected to be seen after $\sim 300$ years. In case $d$, with many strong repeaters, saturation will not occur in the next thousand years, and a steadily increasing repeat rate is expected.

\subsection{z--DM distribution}
\label{sec:zdmdist}

\begin{figure*}
    \centering
    \begin{overpic}[width=0.49\columnwidth]{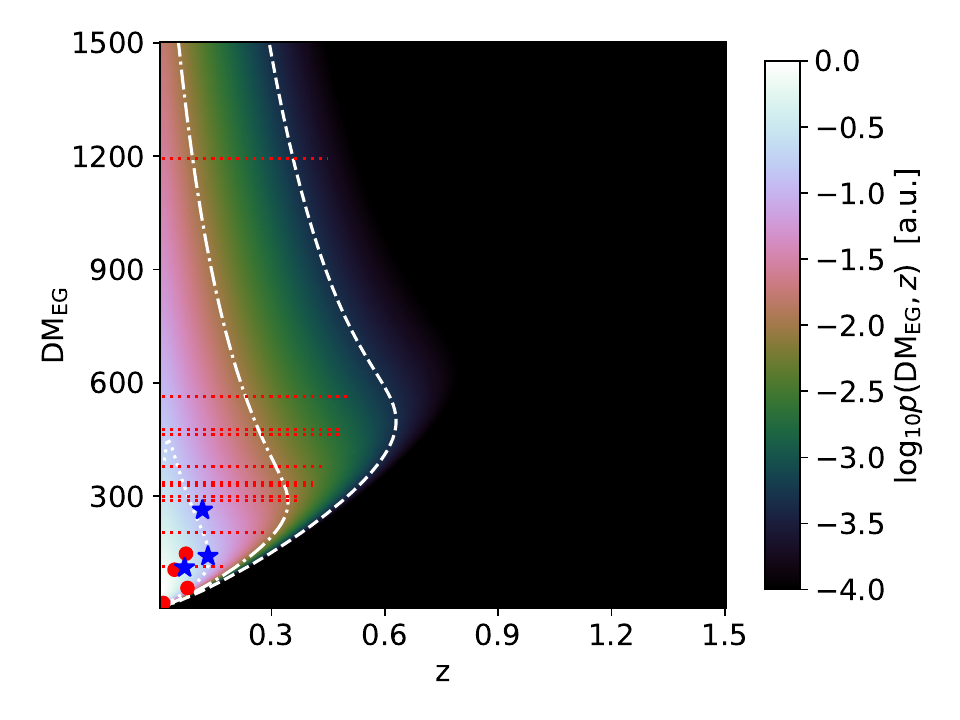}
 \put (0,70) {\large (a)}
\end{overpic}
    \begin{overpic}[width=0.49\columnwidth]{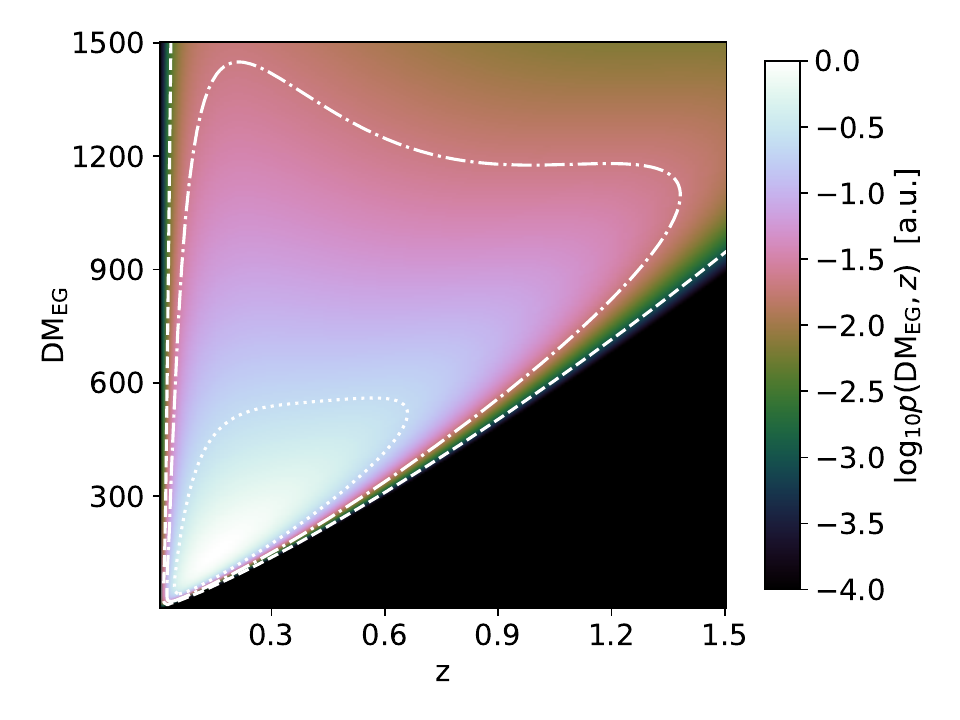} \put (0,70) {\large (b)}
    \end{overpic} \\
    \begin{overpic}[width=0.49\columnwidth]{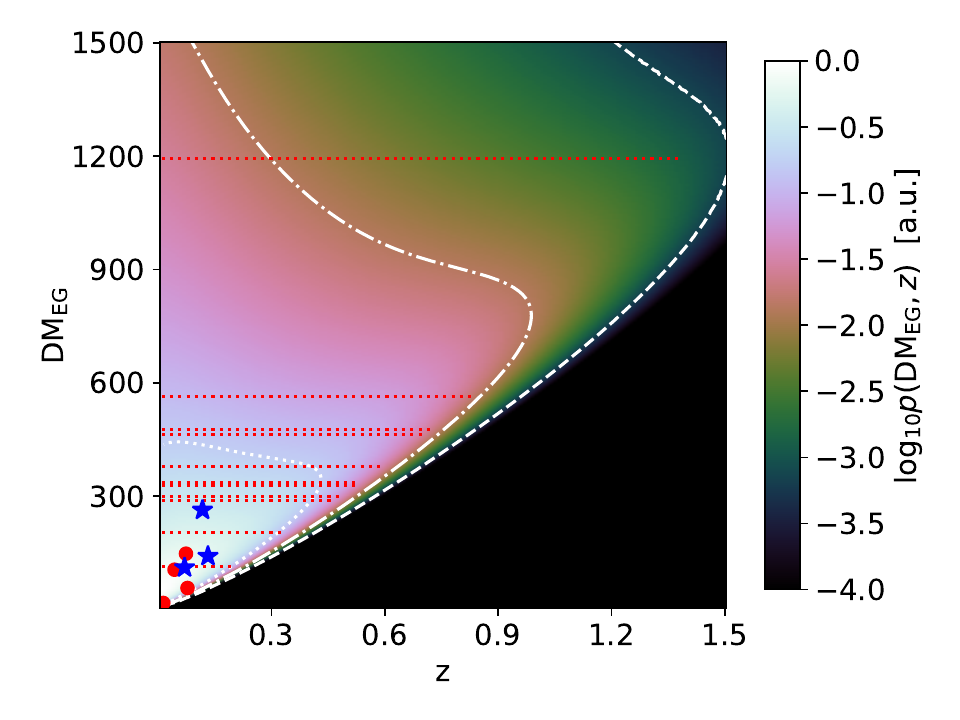} \put (0,70) {\large (c)} \end{overpic}
    \begin{overpic}[width=0.49\columnwidth]{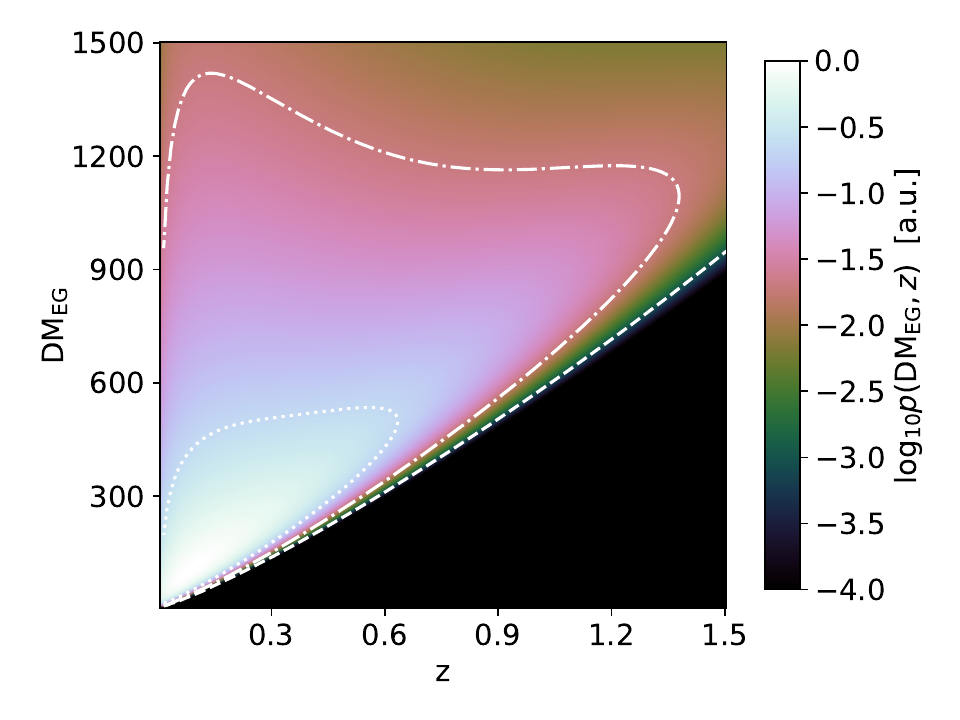} \put (0,70) {\large (d)} \end{overpic}
    \caption{Predicted z--DM distribution of left: repeating and right: single FRBs for cases $b$ (top) and $d$ (bottom). Contours enclose 50\% (dotted), 90\% (dot-dash), and 99\% (dashed) of the probability space. Repeating CHIME FRBs identified in \cat\ with host galaxies are shown as red circles; repeater hosts not from \cat\ are shown as blue stars; and repeating FRBs with no firm host association from \cat\ have their 0-99\% probable redshift range indicated with red lines at their known \dmeg, calculated assuming the  \citet{Cordes_McLaughlin_2003} model for \dmism, and a value of $50$\,\pccc\ for \dmhalo.}
    \label{fig:zdm_repeaters}
\end{figure*}

Only six repeating CHIME FRBs have been localised to their host galaxies \citep{MarcoteRepeaterLocalisation2020,CHIME_M81_2021,Bhardwaj_NGC3252,Fong2021,Ibik2023CHIMEhosts}, though three more associations are highly likely \citep{CHIME2022_rep_loc,Ibik2023CHIMEhosts}; and only one once-off CHIME FRB has a tentative host association \citep{Panther2022_GWFRBhost}. Of these repeaters, four were identified as such in \cat. Furthermore, these have only been localised either because they are nearby, and hence CHIME's angular resolution --- effectively enhanced when using multiple bursts \citep{CHIME_2022_13loc} --- is sufficient to identify the host; or because they repeat rapidly, allowing follow-up observations with arrays with a better angular resolution to identify the host. Thus these represent a highly biased sample, and are unsuited to fitting to data. Nonetheless, the full z--DM distribution of repeating FRBs observed by CHIME can be predicted by the models. These distributions are given in \figref{fig:zdm_repeaters} for cases $b$ and $d$, and are compared to the distribution of singly detected FRBs.

\figref{fig:zdm_repeaters} illustrates how the large tail of the DM distribution arises: it is almost entirely from objects lying well above the Macquart relation. Since repeating FRBs tend to only be detected as such in the nearby Universe, those repeating FRBs lying on the Macquart relation have a smaller range of DMs --- using the 90\% contours, up to $\sim400$\,\pccc in case $b$, and $\sim800$\,\pccc\ in case $d$. Thus the high-DM tail of low-z FRBs doesn't become over-ridden by the larger number of FRBs lying on the Macquart relation in the more-distant Universe, as it is for singly detected FRBs.

This closer proximity of repeating FRBs means that the reduced \dmeg\ of the CHIME repeater sample \citep[$436 \pm 49$ \pccc\ for repeaters, $597 \pm 24$ \pccc\ for apparently once-off bursts;][]{CHIME_2023_25reps}, which has been suggested to be evidence for two populations \citep{2023NatAs...7..374W}, is entirely consistent with expectations from the models. These predict mean repeater \dmeg\ values in the range 460--540 \pccc, and mean single DMs in the range 640--660\,\pccc. While the mean DMs of both samples are slightly over-predicted, the difference is a very good match with expectations.

Even more importantly for future observations, models $b$ and $d$ predict different $z$ distributions. Case $b$ has more low-rate repeaters, from which repeat bursts are only likely at low $z$; while case $d$ has a significant population of high-rate repeaters, which can be detected as repeaters from the more-distant Universe.
The z--DM values of repeaters with likely or confirmed host galaxies matches observations for both cases, but this has only been probed in the low-z, low-DM region. However, case $d$ predicts redshifts will continue to increase with \dmeg, while case $b$ predicts $z \lesssim 0.5$ for all repeaters.
Thus, if a large fraction of repeating CHIME FRBs could be localised, this would enable much more powerful tests of the repeating FRB population.

\begin{figure}
    \centering
    \includegraphics[width=\columnwidth]{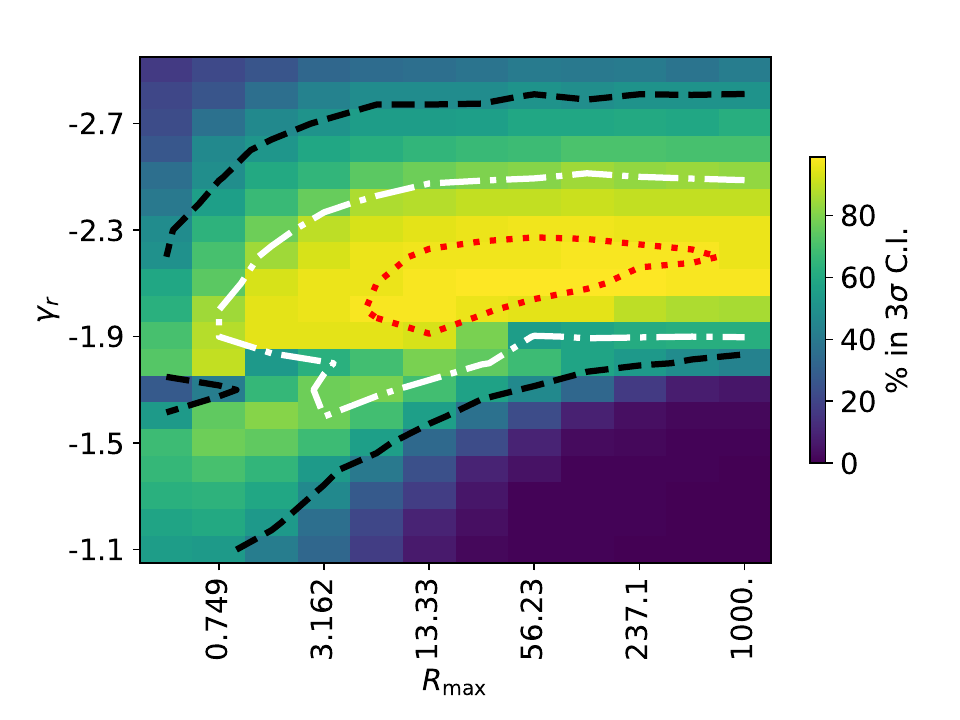}
    \caption{Fraction of Monte Carlo iterations in which trial values of \rr, \rmax\ fall within the $3\sigma$ confidence interval (C.I.), using a sample of simulated repeating FRBs with truth values $\rr=-2.$, $\rmax=31.62$. The contours correspond to regions that fall within the $1$ (red, dotted) 2 (white, dash-dot) and 3 (black, dashed) confidence intervals 50\% of the time.}
    \label{fig:mc_intervals}
\end{figure}

I illustrate this using a toy example, using simulated true values $\rr=-2.2$, $\rmax=30$, and 100 Monte Carlo instances of repeating FRBs from \cat. All FRBs detected as repeaters are assumed to be localised to their host galaxies, yielding their correct $z$ and DM values. For each Monte Carlo sample, the likelihood $p(z,{\rm DM})$ is calculated for all values on the \rr, \rmax\ grid. I do not calculate $p(N_{\rm reps})$, i.e.\ only the position in $z$, DM is accounted for, not the number of repeating FRBs or bursts per repeater. Bayesian $1$-, $2$-, and $3$-$\sigma$ confidence intervals are then constructed for each sample, and the number of MC iterations in which any given value lies in each interval is counted. The result is shown in \figref{fig:mc_intervals}, which shows the expected confidence intervals at each of the three levels. This shows the power of being able to localise repeating FRBs: if all \cat\ repeating FRBs could be localised, the expected $1 \sigma$ accuracy on \rr\ would be $\pm 0.2$, and $\rmax$ would be determined to within a factor of $\sim 10$.

\subsection{Predicted effects on other instruments}
\label{sec:predictions}

I now use cases $b$ and $d$ to estimate the relative rates of single and repeat observations for a sample of other FRB-hunting instruments.
Four systems are considered: ASKAP, in Fly's Eye (FE), incoherent sum (ICS), and coherent (CRACO) mode at 1.3\,GHz; and the Five-hundred-meter Aperture Spherical Telescope (FAST). ASKAP/FE and ASKAP/ICS are
modelled as per \citet{James2022Meth}, while the model of the CRAFT Coherent Upgrade (CRACO) system is described in \citet{James2022_H0}. The parameters for FAST FRB searches are taken from \citet{Niu2021_FAST}, namely a detection threshold of 0.0146\,Jy\,ms for a 1\,ms pulse width at a central frequency of 1.25\,GHz, and time- and frequency-resolutions of $196.608~\upmu s$ and 0.122\,MHz respectively. The FAST receiver is a 19-beam multibeam \citep{Li2018_FAST_Survey}, similarly designed to the 13-beam Parkes multibeam \citep{staveley1996parkesMultibeam}. I therefore take the inverse beamshape $\Omega_b$ used for Parkes, scale up by the ratio of the number of beams ($19/13 \approx 1.46$), and down by the ratio of effective collecting areas ($64^2/300^2 \approx 0.0456$).

\begin{figure*}
    \centering
    \begin{overpic}[width=0.49\columnwidth]{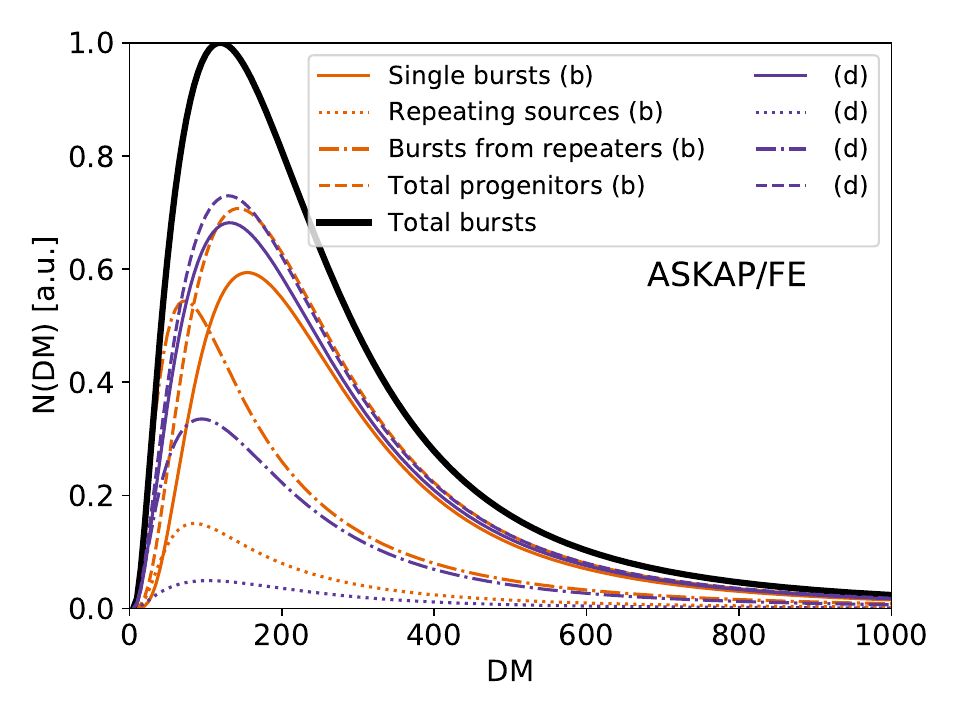}\put (0,70) {\large (a)} \end{overpic} 
    \begin{overpic}[width=0.49\columnwidth]{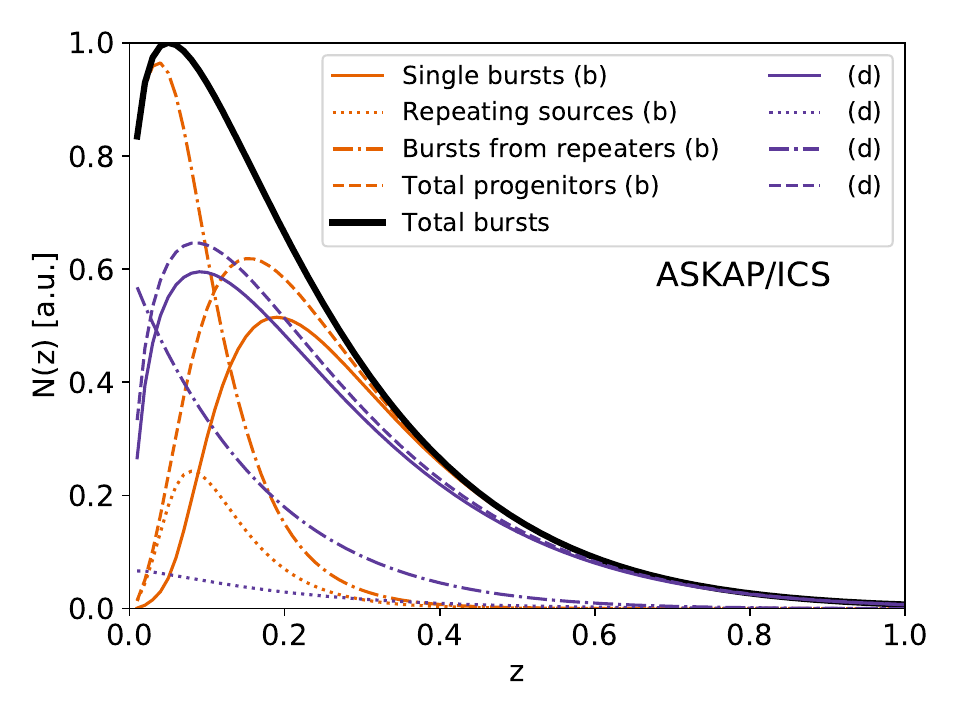} \put (0,70) {\large (b)} \end{overpic}\\
    \begin{overpic}[width=0.49\columnwidth]{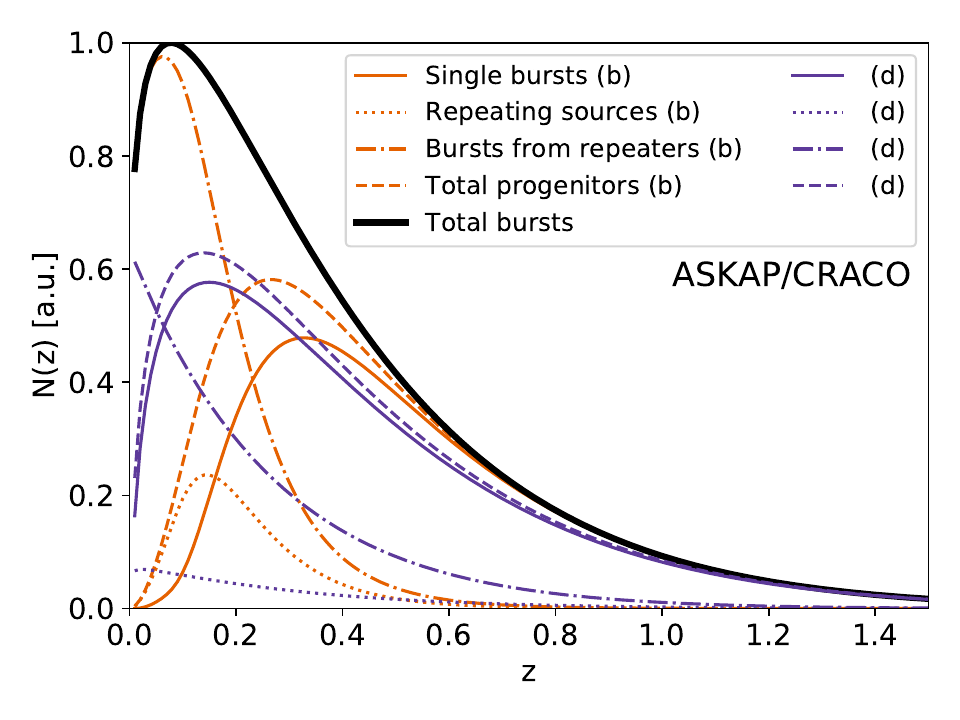} \put (0,70) {\large (c)} \end{overpic}
    \begin{overpic}[width=0.49\columnwidth]{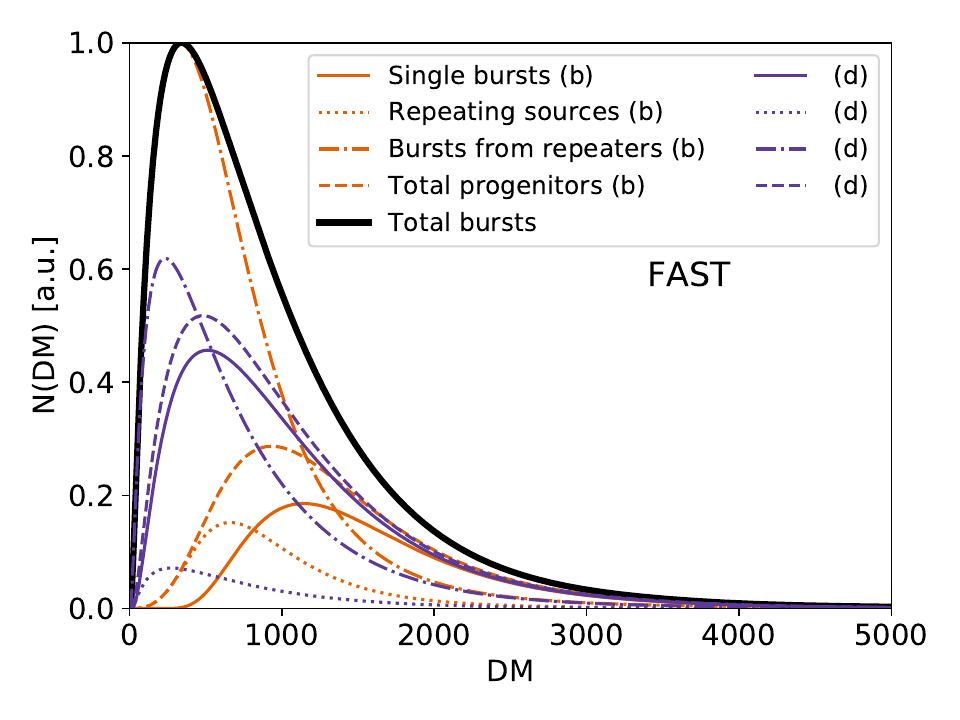} \put (0,70) {\large (d)} \end{overpic}
    \caption{Predictions of the z or DM distributions of repeating FRBs for a selection of past and future FRB surveys for their longest pointing times (see \secref{sec:predictions}), for cases (b) and (d). Shown are the distributions of those repeating FRBs detected as single bursts, as repeaters, the total progenitor distributions, and the total burst distributions, as per \figref{fig:ics_example}.}
    \label{fig:predictions}
\end{figure*}

All these instruments have searched for FRBs with different dwell times. Here, I consider the longest time spent on any given field for each instrument, which will prove most sensitive to the repeating FRB population: 1338.9\,hr for ASKAP/FE \citep{James2020a_followup}, 879.1\,hr for ASKAP/ICS to the end of 2022 (Shannon et al, in prep.), and for ASKAP/CRACO predictions, the expected on-source time of 800\,hr for each of the Deep Investigation of Neutral Gas Origins \citep[DINGO;][see also \url{https://dingo-survey.org/}]{DINGO2023} fields is used. 
For FAST, it is 59.5\,hr when performing follow-up observations on FRB~20121102A \citep{Li2021_FAST_121102}. The normalised estimates are given in Figure~\ref{fig:predictions}, as a function of DM for surveys with poor localisations, and as a function of $z$ for those that typically identify host galaxies.

Qualitatively, all predictions are very similar. The total number of single bursts ranges from 27\% of the total burst distribution (FAST, case $b$) to 70\% (ASKAP/FE, case $d$). The difference between case $b$, which models a repeating FRB population spread over a narrow repetition rate, and case $d$, with a very broad distribution of rates, is marked, predicting 16\% (case $b$) and 6\% (case $d$) of FRB progenitors to repeat for all ASKAP models, and 31\% and 10\% for cases $b$ and $d$ for FAST.

The deficit between total burst number and total progenitors in the low-DM range for ASKAP/FE is not sufficient however to explain the observed deficit that has been previously noted by \citet{James2022Meth}, especially when accounting for the average pointing time for that survey being less than that modelled here. Thus I conclude this effect --- which originally motivated this work --- is most likely a statistical fluctuation.

\citet{Meertrap_2023_sample} have noted that the FAST FRB rate is much lower than predicted. Here, the progenitor rate is predicted to be 40--60\% of the burst rate, which certainly accounts for some, but not all, of the deficit. However, this would have no influence on the observations in drift-scan mode reported by \citet{Niu2021_FAST} due to the very short dwell times ($\sim$13\,s). Thus this deficit must have some other explanation.

A single FRB (20220531A; Shannon et al., in prep) has been discovered in the ASKAP field with 879.1\,hr of observations, against a mean ASKAP detection rate of 350\,hr/FRB. This could simply be a Poissonian under-fluctuation (p-value of 0.285 on a one-sided test), but repetition offers a partial explanation, which would reduce the expected number of progenitors from 2.5 to 1.75--1.86.

Overall, I expect that correct modelling of repeating FRBs will be important for these observations to account for repeater bias in the observed z--DM distribution.

%% file: 08_systematics.tex
\section{Discussion of systematic effects}
\label{sec:systematics}

\subsection{Non-Poissonian repetition}
\label{sec:nonpoissonian}

All FRBs with sufficiently many detected bursts to allow studies of their repetition rates show non-Poissonian behaviour. On timescales of order seconds to hours, bursts from repeaters tend to be clustered \citep[e.g.][]{Gajjaretal2018,2021ApJ...920L..23Z,2023MNRAS.520.2281N}, in a process which is often modelled as a Weibull distribution \citep{Oppermannetal2018}. On longer timescales ($\sim$16--160 days), two repeating FRBs appear to have activity cycles \citep{Chime2020periodicity,2020_121102_periodicity1}, with evidence for frequency dependence in the timing of the windows \citep{2021Natur.596..505P}. Other behaviours include a rapidly increasing/decreasing event rate \citep{FAST2022_active_episode_2}, or `turning on' despite several years of monitoring \citep[see the time-dependence of bursts in][]{CHIME_2023_25reps}. What the true underlying nature of the time-distribution of repeat rates of FRBs is is still under debate --- what is sure is that they are most certainly not Poissonian.

Thus it should be asked: what effect does this have on the modelling? On sufficiently long timescales, FRBs will become inactive, and new repeating FRBs will be born. Therefore, the repeating FRB population studied here can only refer to those FRBs which have been active during the approximate year (three years) corresponding to the CHIME \cat\ (\gold) samples. However, FRB~20121102A has now been studied for over a decade since its first detection \citep{Spitler2014}, and while its properties (DM, RM etc.) do vary \citep{Michilli2018_121102}, no evidence of a systematically reducing rate has been published. Therefore, these considerations likely won't be relevant to current or near-future studies.

Of more relevance are FRBs with inactive windows comparable to, or longer than, the current survey. \citet{CHIME_2023_25reps} shows that at least three FRBs --- 20201130A, 20200929C, and 20201124A --- have numerous bursts in the latter six months studied, but none in the first two years. While some of this may be reflective of a changing search sensitivity and analysis methods (it would be useful for CHIME to release a time-dependent sensitivity to account for this), it is also suggestive that these objects have very long inactive phases. By generating either many bursts or none, a larger population of such objects would mimic a flatter value of \rr\ than the true long-term rate, with repeaters being found at larger distances / DM values. Conversely, for a fixed observation, the fitted value of \rr\ will be flatter than the true value. Since the observed DM distribution of repeaters already favours $\FC \ge 0.5$ and $\rr \le -1.4$, allowing for such behaviour would constrain \FC\ to higher, and \rr\ to lower, values. That is, the limits from this work are sensitive to activity windows on yearly timescales. Activity windows significantly shorter than a year however will have no consequence, since CHIME's coverage is spread uniformly in time, unless the period of these windows lies extremely close to a sidereal day.

Finally, the effect of bursty behaviour in general is to reduce the number of singly detected FRBs, and increase the likelihood of viewing zero or many bursts. The effect of time correlation in bursts will be most pronounced when observations occur all in one block --- when observations are individually very short and spaced far apart in time, any intrinsically bursty distribution will exhibit a Poissonian distribution of burst numbers.

\begin{figure}
    \centering
    \includegraphics[width=\columnwidth]{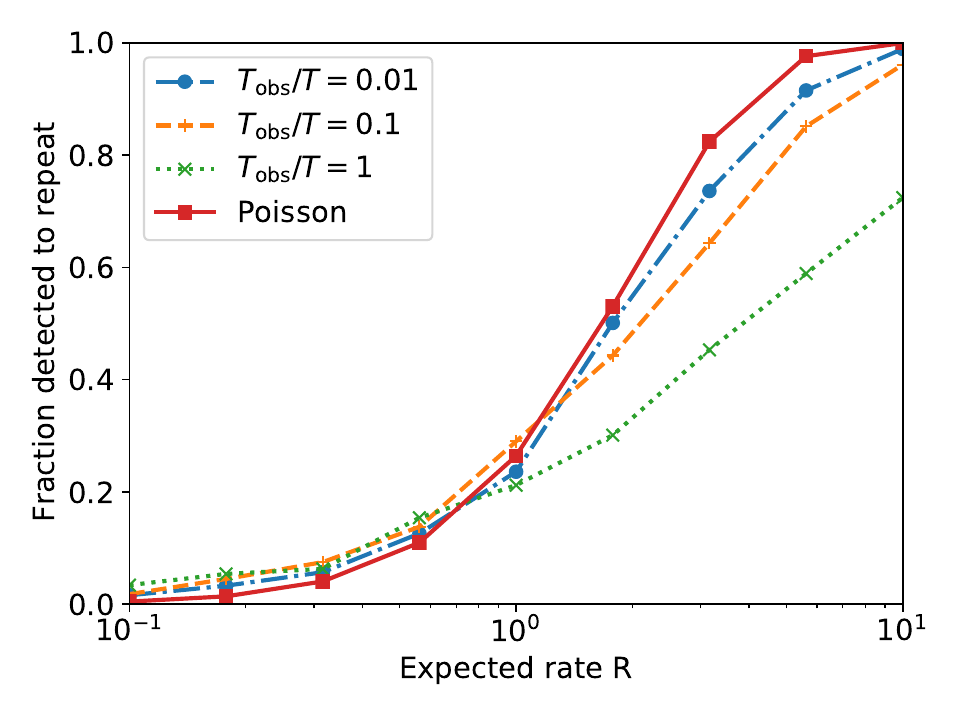}
    \caption{Probability of an FRB being detected as a repeater given its true expected rate $R$, as a function of the fractional observation time $f_{\rm obs}=T_{\rm obs}/T$, for a Weibull burst time distribution with shape index $k=0.34$; and for a Poissonian distribution.}
    \label{fig:weibull_sim}
\end{figure}

To gauge the impact of this effect, I simulate a Weibull distribution of arrival times, using a shape parameter $k=0.34$, as found for FRB~20121102A by \citet{Oppermannetal2018}. CHIME detections are simulated at three declinations: near the North Celestial Pole (NCP), with sources observed continuously ($f_{\rm obs}=1$); approximately $7^{\circ}$ away from the NCP, where sources will be observed a fraction $f_{\rm obs}=10\%$ of the time; and $\sim 30^{\circ}$ from the NCP, which is modelled as a source being observed $f_{\rm obs}=1\%$ of the time. The expected repetition rates are modelled relative to the total time on-source in a calendar year, and this rate $R_{\rm obs}$ is varied from 0.1 to 10. I simulate 1000 Weibull sequences over 365 sidereal days, beginning each sequence well before the start of the year to ensure the sequence start time does not influence results. If a burst occurs in the first $f_{\rm obs}$ fraction of a day, it is counted as detected. The number of simulations resulting in none, one, or multiple detections are recorded.

The results are given in \figref{fig:weibull_sim}. As predicted, the fraction of FRBs detected to repeat twice or more is greater than Poissonian for low expected rates, and less for high expected rates. However, only very near the NCP is this effect large, where a deficit of $\sim 30\%$ of repeating FRBs relative to Poisson rates are found.

A more accurate estimate of the effect of burstiness on these results however is the ratio of single to repeat bursts. Weighting the results by $R^{-2}$, to represent there being less rapidly repeating FRBs than rarely repeating ones, $f_{\rm obs} = 0.01$, $0.1$, and $1$ respectively produces 30\%, 120\%, and 310\% more repeating FRBs relative to single bursts than expected for a Poissonian distribution. This effect would also alter the $z$--DM distribution of repeating FRBs, with bursty distributions favouring discoveries in the more distant Universe; and flatten the distribution of observed burst rates, undermining the fitting of \secref{sec:fit_rates}.
It is quite possible that the excess of repeating FRBs in \cat\ at high declinations, while not statistically significant,  is due to this effect.

Nonetheless, I do not wish to over-emphasise this effect. For the majority of the sky seen by CHIME, the increase in observed repeaters due to burstiness with $k=0.34$ is tens of percent. Furthermore, this choice of $k=0.34$ is an extreme example --- analysis of FRB~20121102A has shown that at timescales of seconds to hours, the wait-time distribution is close the Poissonian, with a best-fit Weibull index of $k=0.79$ \citep{2023MNRAS.520.2281N}.

Given the complexity of the issue, I defer an analysis of repetition that includes bursty behaviour to a future work --- which should also include a fit of such behaviour to CHIME arrival-time data.

\subsection{Time-frequency structure of bursts}
\label{sec:width}

CHIME has shown that FRBs identified to repeat tend to have bursts which are broader in time by a factor of approximately two, with complex time-frequency structure \citep{CHIME_catalog1_2021,CHIME_morphology_2021}. Whether this is due to two intrinsically different FRB populations, or a smooth transition in the properties of one population \citep[e.g.\ an aging effect whereby less active --- and presumably older --- objects have modified emission physics, perhaps due to a change in beaming angle, as suggested by ][]{Connor2021_beaming} is still up for debate. What is clearly true is that the increased time-width of FRBs more likely to be detected as repeating will make them harder to detect for a given fluence. This effect is not included in the current work.

There are two methods of analysing this effect. The first is to modify the simulated burst width to increase with FRB repetition rate. This will act to suppress the number of repeating FRBs in the $z \lesssim 1$
range where intrinsic FRB width, rather than dispersion smearing, dominates the apparant FRB width, and hence sensitivity. The result will be that more active repeaters --- which are in any case only observed as such at low $z$ --- become less detectable. Including this effect would require direct use of the CHIME pulse injection sample, since the published efficiency function analysed in \secref{sec:sensitivity} has already been averaged over the burst width distribution.

The second method to tackle this problem is to note that repetition rate $R$ can be thought of as an effective repetition rate. After all, $R$ can only ever be defined as the rate above a given energy threshold (here, $10^{39}$ erg), which must also be coupled to some assumption about the time--frequency properties of those bursts which affects their detectability. If more-strongly repeating FRBs tend to produce wider bursts, this reduces their detectability, and hence their apparent rate will decrease. In other words, this will cause \rr\ to steepen slightly from its true value. In the regime where detectability is limited by intrinsic width $w$, \snr$\sim w^{0.5}$, and hence for a burst luminosity function with cumulative slope $\gamma = -1$, the detection rate will reduce as $w^{-0.5}$. If the width $w$ scales linearly with the intrinsic rate $R$ (and the real effect is unlikely to be this strong), this would then steepen the apparent value of \rr\ by an extra factor of $-0.5$.

Since either method requires an accurate estimate of the relationship between intrinsic FRB rate and measured width $w$, and this is not currently possible, I consider the second method appropriate, which means that a little care must be taken in interpreting the estimate of \rr\ found in this work.

\subsection{Systematic effects in CHIME data}
\label{sec:gold}

I conclude the discussion of systematics with an analysis of the data being used. As noted in \secref{sec:data}, the threshold for identifying a repeat burst in \cat\ was lower than for an initial burst. The best way of removing this effect would be for CHIME to publish an FRB catalog with those repeat bursts that passed threshold only because they were repeaters removed or otherwise identified as such. The dependence on \snr\ should be overcome by use of the CHIME pulse injection data \citep{chime_injection_2022}, or a publication of the parameterised multi-dimensional selection function used in \citet{CHIME_catalog1_2021}. It has been checked that excluding FRBs with high Galactic DMs does not significantly change the DM fits, but future analyses with \zdm\ should in any case include an uncertainty term for this contribution. Overall, I do not think that this analysis is currently constrained by such systematic effects.

I have been reluctant to make quantitative comparisons in this work with CHIME's new \gold\ sample of 25 repeating FRBs however. There are several reasons why interpretation is difficult.
Firstly, note that this work is in fact a discovery of $34$ new repeating FRBs, given that the ``gold'' sample of 25 sources has an estimated contamination of 0.5 from coincidences between two or more unrelated FRBs, and the authors also publish a ``silver'' sample of 14 repeaters, with a total contamination rate in the combined gold and silver samples of 5. It is unknown whether or not CHIME have detected two or more bursts from a true repeater which has a higher contamination fraction, and is thus not included in any sample. This measurement therefore has a $5^{0.5} = 2.2$ systematic uncertainty reflecting the uncertainty in the expected contamination, and a $34^{0.5} = 5.8$ statistical deviation from the expected mean number of repeater discoveries, for a total uncertainty of $(5+34)^{0.5} = 6.2$, or 18\%.

The time period used for the search, from September $30^{\rm th}$ 2019 to May $1^{\rm st}$ 2021 (579 days), is approximately 70\% longer than, and does not overlap with, the period of July $25^{\rm th}$ 2018 to July $1^{\rm st}$ 2019 used for \cat\ (342 days) where 16 repeating FRBs were identified. Hence, the detection rate has increased by a factor of $1.26 \pm 0.23$, 
consistent both with the predictions of an increasing discovery rate in \secref{sec:tdist}, and with a constant rate. Aside from Poisson error, this increase could also be due to a higher efficiency of CHIME data-taking, or a lower threshold for including bursts in the analysis. This question should be revisited once it becomes possible to normalise the two samples, for an accurate rate comparison.

Secondly, the identification of repeating FRBs in \gold\ placed a cut on the chance coincidence probability. This cut is strictest where the rate of FRBs in DM--$\delta$ space is highest: at high declinations, and intermediate DMs. Therefore, the gold sample of 25 repeating FRBs is biased towards very low or high DMs at low declinations. Conversely, the silver sample may have the opposite bias, both because repeaters at high declinations and intermediate DMs will preferentially be placed there, and because this region has more chance coincidences. \citet{CHIME_2023_25reps} do not publish estimates of the contamination probability in $\delta$-DM space, which would be required to account for this effect.

%% file: 09_conclusions.tex
\section{Comparison with literature results}
\label{sec:askap}

\subsection{ASKAP follow-up observations}

The only other result of which I am aware that has limited \rmin, \rmax, \rr\ is \citet{James2020b_popreps}. 
Those authors used the observation of repetition in only one of 27 FRBs detected by ASKAP \citep{Kumar2019,James2020a_followup} to constrain $\rr < -1.94$ (those authors use $\zeta$ for $\rr$) and $\rmin < 10^{-2.9}$\,day$^{-1}$. Limits on $\rmax$ are not published, though values of $\rmax \le 100$ are investigated.

The values of $\rmin$ and $\rmax$ used in that work are applicable to $1.3$\,GHz observations, and are measured relative to an energy threshold of $10^{38}$\,erg. I therefore scale to a threshold of $10^{39}$\,erg by reducing those rates by a factor of $10^\gamma \approx 0.13$, and to the mean CHIME frequency of 600\,MHz by increasing the rates by a factor of $(600/1300)^\alpha \approx 4.4$, for a total adjustment factor of $0.55$.

\begin{figure}
    \centering
    \includegraphics[width=\columnwidth]{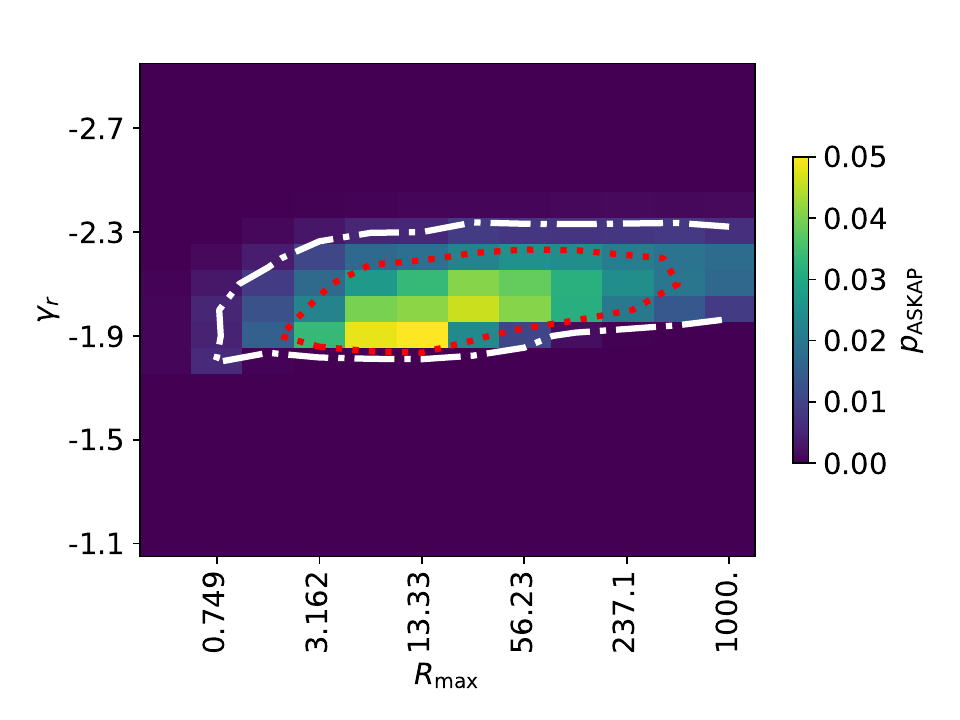}
    \caption{Bayesian posterior likelihoods, $p_{\rm ASKAP}$, from follow-up observations of ASKAP repeaters \citep{James2020b_popreps} for the parameter values investigated here.}
    \label{fig:askap}
\end{figure}

I extract the likelihoods from that work at the values of \rmin, \rmax, and \rr\ used here, and plot the Bayesian posterior in \figref{fig:askap}. The best-fit region agrees remarkably with results from CHIME FRB data, but has a much tighter constraint on \rr. That two very different measurements with very different instruments converge to the same properties of the repeating FRB population is strong evidence that this model is a reasonable approximation to the underlying truth. This has implications for FRB progenitor models, as discussed in \citet{James2020b_popreps}.

The one major discrepancy between the results which is hidden by \figref{fig:askap} however is that the best-fit values of \rmin\ found by \citet{James2020b_popreps} are a factor of $\sim 100$ lower than that found for CHIME. Equivalently, the results of \citet{James2020b_popreps} would under-predict the number of repeating CHIME FRBs. There may be several causes for this.

Firstly: the results for \citet{James2020b_popreps} assumed a Weibull distribution with $k=0.34$. As discussed in \secref{sec:nonpoissonian}, a bursty distribution requires less-rapid repeaters to produce the same number of repeating FRBs, and could therefore allow the values of \rmin\ found here to be lower, and more consistent with the ASKAP results. Secondly: the results here are only weakly constraining on \rmin, and only the probabilities at the best-fit values have been used, rather than marginalising over \rmin. Thirdly: repetition behaviour at 1.3\,GHz and 600\,MHz might be more different than the simple scaling above would suggest. A fuller investigation will require an improved model of FRB time--frequency structure. Fourthly: the limits from \citet{James2020b_popreps} used non-localised repeaters, with conservatively large distance estimates from the Macquart relation assuming no host contribution, thus introducing a bias in those results (albeit one which would push \rmin\ and \rmax\ to lower values). Fifthly and finally: not all FRBs may repeat. The upper limit on \rmin\ found by \citet{James2020b_popreps} is driven primarily by the lack of observed repetition from FRB~20171020A. If this FRB is intrinsically a once-off event, then those results weaken significantly, allowing higher values of \rmin.

The current implementation of repetition in the \zdm\ code does not allow for easy estimation of FRB repetition parameters from follow-up observations.
It would be useful to develop such a method to allow the results of follow-up observations to also be fit in a self-consistent manner.

\subsection{Absolute number of repeaters and persistent radio sources}
\label{sec:numbers}

\begin{figure}
    \centering
    \includegraphics[width=1.05\columnwidth,trim={0.5cm 0.2cm 0 0.8cm},clip]{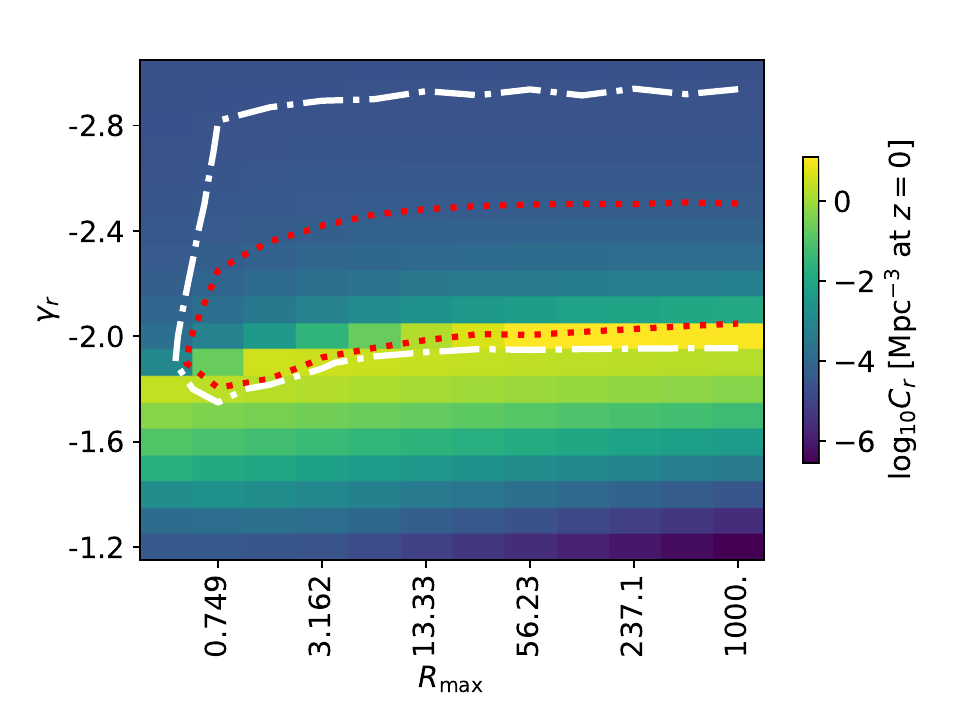}
    \caption{FRB progenitor population density $C_r$, as a function of \rmax\ and \rr\, with 68\% and 95\% contours from \figref{fig:ptot} overplotted.}
    \label{fig:Cr}
\end{figure}

The total number of repeating FRBs, $C_r$, is a function of their rate of birth and active lifetime. Estimates of their birth rate vary greatly according to their progenitor model, and range from $10^5$\,Gpc$^{-3}$\,yr$^{-1}$ for core-collapse supernova \citep{2014ApJ...792..135T}, to 0.02\,Gpc$^{-3}$\,yr$^{-1}$ for NS-NS binary mergers in globular clusters \citep{2020ApJ...888L..10Y}. Their lifetime is also unknown: while magnetar fields are expected to decay on timescales of order $10^4$\,yr \citep{2000ApJ...529L..29C}, precisely how this relates to their time as an active emitter of FRBs will depend on the microphysics of emission.

\figref{fig:Cr} plots the implied value of $C_r$ as a function of \rmax\ and \rr, compared to limits on those parameters from \figref{fig:ptot}. By far the largest number of FRB progenitors are found near $\rr=-2$. At flatter (less negative) values of \rr, the repeater distribution is dominated by so many rapidly repeating FRBs that very few are required, while at steep (more negative) values, \rmin\ must be relatively high (and hence $C_r$ not too large) to prevent singly detected objects overwhelming the total FRB rate. That this region is just allowed within the current limits means that the total uncertainty on $C_r$ is huge, ranging approximately between $10^{-5}$ and $10$\,Mpc$^{-3}$ at $z=0$. Assuming a lifetime of $10^{4}$\,yr, this corresponds to birth rates of $1$--$10^6$\,Gpc$^{-3}$\,yr$^{-1}$, which is compatible with most FRB progenitor models.

The total number of strong repeating FRBs is also of interest. Aside from their obvious identification via detection of their bursts, these objects might be identified via their association with persistent radio sources (PRS), which can be identified in radio surveys. There is some evidence that PRS are more likely to be associated with strong repeaters \citep{Chatterjee2017,Niu2022}, though this evidence is not conclusive \citep{Law2022_PRS}.

In this context, \citet{Law2022_PRS} estimate the total number density, and differential power-law slope, of the observed repeating FRB population using CHIME data, as in \secref{sec:fit_rates}. Assuming all CHIME FRBs come from intrinsic repeaters, they use the observed distribution of the number of bursts to have a power-law slope of $=-1.5$, calculated via the maximum-likelihood estimator of \citet{Crawford1970}. This method thus determines the slope of the `source counts' distribution of repeaters, which is largely insensitive to the intrinsic value of \rr\ (see \figref{fig:mc_histogram}). They also estimate the mean number of observed repeat bursts per source, $\left< R_{\rm obs} \right>=1.2$\,day$^{-1}$, assuming a minimum repetition rate equal to the inverse of $T_{\rm \cat}^{-1}$. Again, this estimate is based on the observed repetition rates, rather than the intrinsic rates, which explains why the derived mean rate is somewhat higher than the values of \rstar\ found here. Their derived number density of repeaters, being between $22$ and $5.2 \cdot 10^3$\,Gpc$^{-3}$ in the no-beaming case, are both more constraining, and on-average lower, than derived here. \citet{Law2022_PRS} estimate that the fraction of FRB sources with associated PRS is between 0.06 and 0.36: the much higher uncertainty on the intrinsic source density found in this work suggests that surveys identifying PRS independently of FRBs might be a better tracer of the intrinsic number of FRB sources.

It should be noted that the estimates above assume isotropic emission. However, to first order, beaming should have no effect. Including beaming means that the intrinsic number of bursts per repeater is greater by $4 \pi / \Omega_{\rm frb}$, where $\Omega_{\rm frb}$ is the solid angle subtended by each FRB. However, so is the intrinsic single burst emission rate. Hence, the implied number of repeaters to reproduce the detected rate is identical, and beaming angle has no effect. Only in the case that $\Omega_{\rm frb}$ varies with repetition rate $R$, as suggested by \citet{Connor2021_beaming}, is this scaling broken.

\section{Conclusions}
\label{sec:conclusions}

I have implemented a model for repeating FRBs in the \zdm\ code, allowing for a power-law distribution of intrinsic FRB repetition rates. A population of repeating FRBs will result in an apparent deficit of progenitors (single and repeat FRBs) in the nearby Universe, with the effect becoming increasingly strong with observation time per pointing. I show that this effect is significant for current observations with ASKAP, FAST, and CHIME, and hence that future FRB population modelling should include a simultaneous fit to FRB repetition parameters as well as the current set of cosmological parameters, host galaxy properties, population evolution, and the luminosity function. Such a fit is computationally infeasible with current methods implemented in the \zdm\ code, and hence nested sampling techniques should be implemented in the future.

I have therefore fit a power-law model of repeating FRB rates, with differential slope \rr\ between rates \rmin\ and \rmax\ (defined as bursts per day above $10^{39}$\,erg, with a Poissonian distribution of arrival times), to CHIME Catalog 1 data. The model of the CHIME experiment includes beamshape, DM response, and declination-dependent exposure. This model can accurately reproduce the distribution of singly detected CHIME FRBs when using FRB population parameters with a steep dependence of FRB rate on frequency (rate$\propto \nu^{-1.91}$), consistent with previous fits to ASKAP and Parkes data at the 90\% confidence level. Holding this parameter set fixed, I find that the distribution of repeating FRBs in DM, $\delta$, and $N_{\rm burst}$ space, as well as their number compared to apparently once-off bursts, is well-reproduced when assuming the entire FRB population is explained by a single population of repeating FRBs with $\rr=-2.2_{-0.8}^{+0.6}$ (68\% C.I.), and $\rmax \ge 0.75$. Limits on \rmin\ are less well-constrained. In particular, results are consistent with the DM deficit of repeating FRBs found by \citet{CHIME_2023_25reps}. This remains the case unless less than 50\% of all CHIME single FRBs are due to intrinsic repeaters, at which point the predicted DMs of CHIME repeaters become too high to be consistent with data.

I also make predictions for the effects of repetition in current and future experiments. Localising CHIME repeating FRBs to obtain their redshifts would provide strong constraints on the repeating FRB population, and I urge that optical follow-up observations preferentially target these sources. Furthermore, I predict that the number of repeating FRBs identified by CHIME will continue to increase for at least the next hundred years' worth of observations, and potentially for a much longer timespan. This is currently consistent with the new sample of repeaters released by \citet{CHIME_2023_25reps}, but systematic effects in that data set make more precise comparisons difficult.

An estimate is made of the systematic effects of correlations between repetition rates and FRB width, non-Poissonian arrival times, and differences between the full CHIME response to FRBs described by \citet{chime_injection_2022} and used by \citet{Shin2022} and the method used here. Thus burstiness may be responsible for the apparent excess of repeating FRBs at high declinations, but that other effects are likely small, though may alter the best-fit value of \rr\ away from its true value. Improved modelling of CHIME may be required in the future, though this current implementation already accounts for most effects described by \citet{CHIME_2023_25reps}.

The most significant outcome is that limits on repetition parameters derived here agree with estimates of repeating FRB parameters produced from follow-up observations to ASKAP FRBs, despite the large differences in detection systems and method of estimation.

In conclusion, I emphasise that while this work is not conclusive evidence for the entire FRB population being explained by a single population of intrinsically repeating progenitors with a broad distribution of repetition rates, it certainly shows that current observations of repeating and single FRBs by CHIME are completely consistent in terms of DM, declination, number of bursts from repeaters, and the relative number of FRBs observed once and multiple times with this scenario.

%% file: 10_appendix.tex
\section{Alternative beamshape modelling}
\label{app:beamshape}

To evaluate the effect of different beam approximations on CHIME's response to repeating FRBs, I use the `strong repeaters' model of \secref{sec:qualitative}, and assume a source at an example declination of $\delta=30^{\circ}$.
Three methods of parameterising the beam are considered. The first, and simplest, is to calculate a time-weighted effective beam sensitivity using
\begin{eqnarray}
    t_{\rm eff}^\prime & = & \sum_i t_i \overline{B}_i^{1.5} \\
\end{eqnarray}
which weights each frequency-averaged beam value, $\overline{B}$, by the cumulative source counts index of $1.5$. This reduces the time spent observing a repeating FRB to a single effective time, $t_{\rm eff}^\prime$, at beam sensitivity $B=1$. However, the characteristic sensitivity, $B_{\rm eff}$, is better calculated by weighting these contributions by $B$, such that
\begin{eqnarray}
 B_{\rm eff} & = &  t_{\rm eff}^{\prime -1} \sum_i \overline{B} t_i \overline{B}_i^{1.5}.
\end{eqnarray}
At that sensitivity, the effective time $t_{\rm eff}$ is greater than that at $B=1$, i.e.\
\begin{eqnarray}
    t_{\rm eff} & = & t_{\rm eff}^\prime B_{\rm eff}^{-1.5}.
\end{eqnarray}
The second method parameterises $T(B)$ by histogramming the frequency-averaged beamshape $T(\overline{B})$, which is that standard method used in this work. The third method first creates a histogram for each frequency, and then averages those, to produce $\overline{T}(B)$. Here, 15 bins in $B$ are used, equally log-spaced from $10^{-3}$ to unity.

\begin{figure}
    \centering
    \includegraphics[width=\columnwidth]{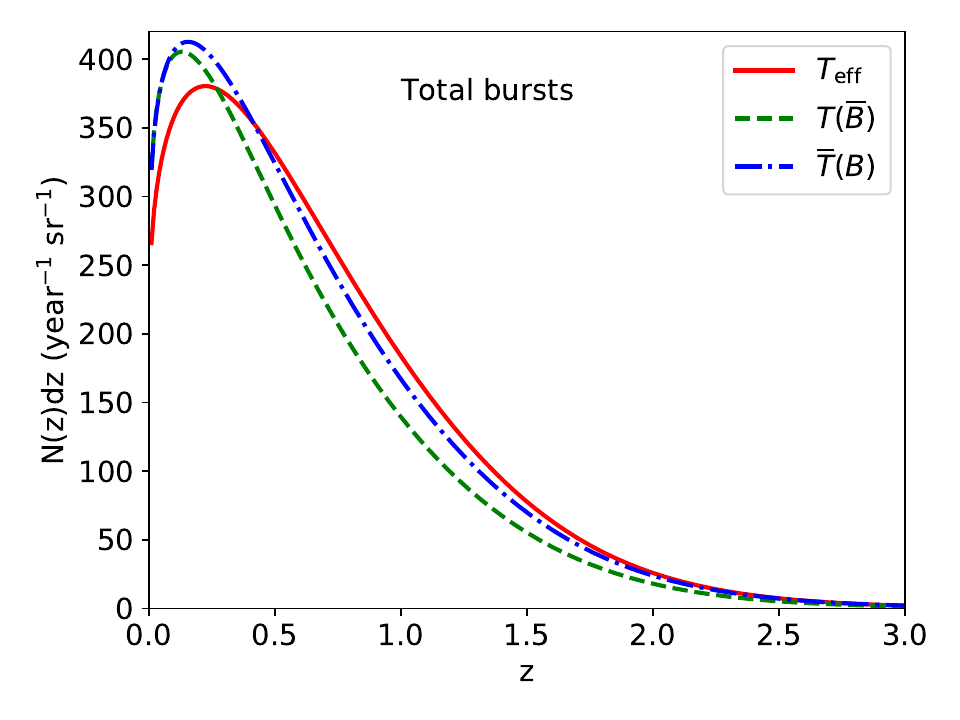}
    \caption{Comparison of the total burst rate for three methods of calculating $T(B)$: a single effective value ($T_{\rm eff}$), using the frequency-averaged beam $T(\overline{B})$, and averaging the time after calculations at each frequency $\overline{T}(B)$.}
    \label{fig:beam_comparison}
\end{figure}

The total expected number of bursts (per year per steradian) is given in \figref{fig:beam_comparison}. Systematic errors due to the discretisation of $T(\overline{B})$ and $\overline{T}(B)$ are less than the line widths in the figure, whereas $T_{\rm eff}$ differs from $\overline{T}(B)$ purely because of discretisation to a single value. The difference between the latter two cases is clearly significant, with the use of a single characteristic value of sensitivity over-predicting the burst rate at high $z$, and underpredicting it at low $z$. As noted in \citet{James2022Meth}, this is simply because using the $T_{\rm eff}$ method (method 1) is an over-simplification.
However, the difference between $T(\overline{B})$ and $\overline{T}(B)$ is more subtle. When bursts are broadband, the instrumental response will be averaged over the total bandwidth, and hence the frequency-averaged sensitivity, $T(\overline{B})$, applies. However, when bursts exhibit limited band occupancy, the sensitivity at a specific frequency range is relevant. The total burst rate would then be better predicted by first calculating the time-average of the sensitivity in each fraction of the total bandwidth, i.e.\ $\overline{T}(B)$. This latter method also leads to more bursts being detected when the slope of the cumulative source-counts (`logN-logS') distribution is steeper than $-1$: in a toy example of a beam with sensitivities $0.1$ and $0.9$ at two different frequencies and Euclidean source counts, the rate using the $T(\overline{B})$ method will be $[0.5(0.1+0.9)]^{1.5}=0.35$, while using the $\overline{T}(B)$ method, it will be $0.5(0.1^{1.5}+0.9^{1.5})=0.41$.

\citet{CHIME_catalog1_2021,CHIME_morphology_2021} have found that the band occupancy varies significantly from burst-to-burst, with some bursts exhibiting a broad-band morphology, and others --- particularly those from observed repeaters --- being band-limited. However, the majority are not significantly band-limited, and thus $T(\overline{B})$ has been adopted as the default beam parameterisation.

\section{Declination binning}
\label{app:decbins}

In order to account for the declination-dependent exposure of CHIME, six bins in declination have been chosen, and the CHIME exposure averaged over each bin. The choice of the number of bins is somewhat arbitrary --- here, I characterise the error made through averaging by comparing predictions made with a much larger number of declination bins.

\begin{figure}
    \centering
    \includegraphics[width=\columnwidth]{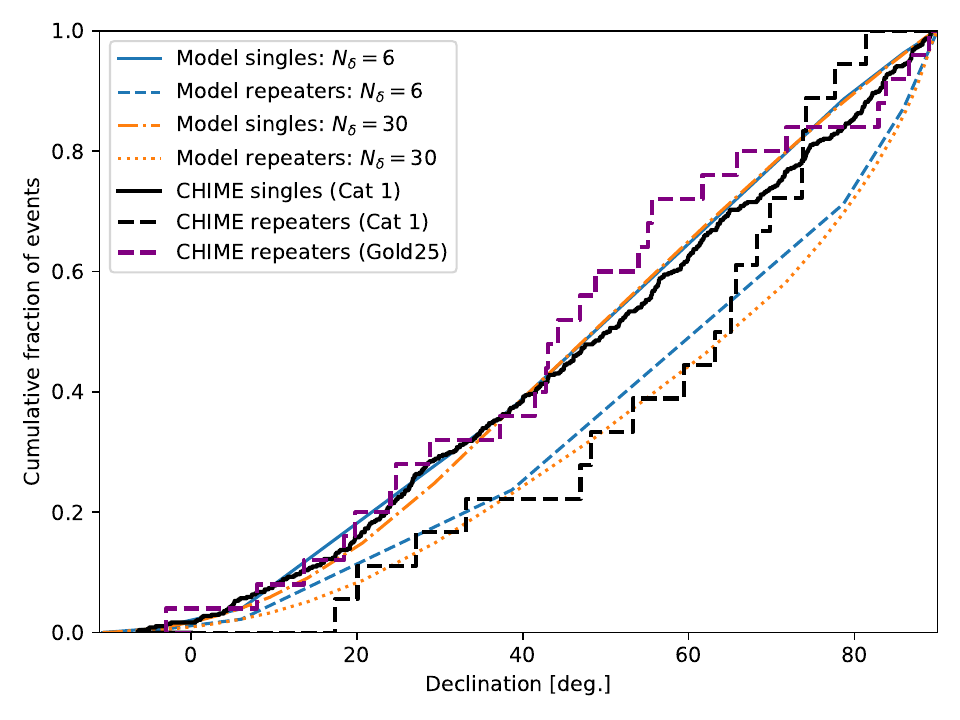}
    \caption{Cumulative distribution of single (solid) and repeat (dashed) FRBs, as predicted from simulations with $N_\delta=6$ (blue) and $N_\delta=30$ (orange), with the `min $\alpha$' parameter set with CHIME's DM selection function. This is compared to CHIME catalog 1 single and repeat bursts (black), and also repeat bursts including the 3-year sample (purple).}
    \label{fig:decbinfig}
\end{figure}

For this test, the `min $\alpha$' model with the CHIME exposure from Table~\ref{tab:extreme_params} is used, and calculations are performed using $N_\delta=6$ and $30$. Predictions for the declination-dependence of the CHIME FRB populations are compared to data in Figure~\ref{fig:decbinfig}. Other models produce qualitatively similar behaviour.

From Figure~\ref{fig:decbinfig}, it can be seen that while $N_\delta = 30$ produces quantitatively different predictions for the $\delta$-dependence of the repeating FRB population, differences between predictions and observations are generally larger, such that using a small $N_\delta$ won't affect the conclusions here. Thus I have used $N_\delta = 30$ only for display purposes in \figref{fig:dec_distribution}.